\documentclass[journal]{IEEEtran}
\usepackage{cite}
\usepackage{amsmath,amssymb,amsfonts}
\usepackage{algorithmic}
\usepackage{graphicx}
\usepackage{textcomp}
\usepackage{amssymb}
\usepackage{graphicx}
\DeclareGraphicsExtensions{.pdf,.png,.jpg}
\usepackage{amsfonts}
\usepackage{mathtools,amssymb}
\usepackage{algorithm}
\usepackage{algorithmic}
\usepackage[utf8]{inputenc}
\usepackage[english]{babel}
\usepackage{upgreek}
\usepackage{cite}
\newtheorem{theorem}{Theorem}

\newtheorem{lemma}{Lemma}
\makeatletter
\def\munderbar#1{\underline{\sbox\tw@{$#1$}\dp\tw@\z@\box\tw@}}
\makeatother
\usepackage{accents}
\usepackage{textcomp}
\usepackage{multirow}
\usepackage[table,xcdraw]{xcolor}
\usepackage{verbatim}
\usepackage{subfig}
\ifCLASSINFOpdf
  
\else

\fi

\begin{document}

\title{Joint Orthogonal Band and Power Allocation for Energy Fairness in WPT System with Nonlinear Logarithmic Energy Harvesting Model}

\author{Jaeseob~Han,~\IEEEmembership{Student Member,~IEEE,}
        Gyeong Ho~Lee,~\IEEEmembership{Student Member,~IEEE,}
        Sangdon~Park,~\IEEEmembership{Member,~IEEE,}
        and~Jun Kyun~Choi,~\IEEEmembership{Senior Member,~IEEE}
\thanks{M. Shell was with the Department
of Electrical and Computer Engineering, Georgia Institute of Technology, Atlanta,
GA, 30332 USA e-mail: (see http://www.michaelshell.org/contact.html).}
\thanks{J. Doe and J. Doe are with Anonymous University.}
\thanks{Manuscript received April 19, 2005; revised August 26, 2015.}}

\markboth{Journal of \LaTeX\ Class Files,~Vol.~14, No.~8, August~2015}%
{Shell \MakeLowercase{\textit{et al.}}: Bare Demo of IEEEtran.cls for IEEE Journals}

\maketitle

\begin{abstract}
Wireless power transmission (WPT) is expected to play an important role in the Internet of Things services by providing the perpetual operation of IoT sensors. However,  to prolong the IoT network's lifetime, the efficient resource allocation algorithm is required, in particular, the energy fairness issue among IoT sensors has been a critical challenge of the WPT system.
In this paper, considering energy fairness as the minimum received energy of all energy poverty IoT sensors (EPISs), we allocate orthogonal frequency bands to several EPISs and transfer the RF power on each orthogonal band, using energy beamforming.  Based on the energy poverty, we propose orthogonal frequency bands assignment rule, granting the priority to the EPISs with less received energy.
We also formulate two transmission power allocation problems, incorporated the nonlinear logarithm-energy harvesting (EH) model. 
First, the total received power maximization (TRPM) problem is presented and solved by combining the well-known Karush-Kuhn-Tucker (KKT) conditions with the modified water-filling algorithm. 
Second, the common received power maximization (CRPM) problem is formulated and the optimal solution is derived using the iterative bisection search method. 
To apply the bisection search method to the problem, this paper proposes a method of specifying the scope of the solution for the objective function defined by the sum of monotonous functions.
In numerical results, assuming the mobility of EPISs by the one-dimensional random walk model, the effectiveness of the mobility of EPISs on the minimum received energy of all EPISs is presented. In addition,  this paper analyzes the difference in performance among the nonlinear EH model and linear EH model in CRPM and TRPM problem. Finally, the performance of the proposed resource allocation schemes is verified by comparing other resources allocation schemes, such as Round robin and equal power distribution
\end{abstract}


\IEEEpeerreviewmaketitle

\section{Introduction}
\IEEEPARstart{W}{ith} the Internet of Things (IoT) emerging as a promising technology for the hyper-connected society, a large number of IoT sensors have been placed in various IoT application domains, such as the smart city and smart home. Consequently, the IoT technology will cover every part of our lives.
However, since most of IoT sensors are battery-powered wireless devices, these devices have an energy poverty problem. To tackle this problem, the concept of radio-frequency (RF) wireless power transmission (WPT) technology has been extensively investigated. Since it can transmit power wirelessly to mobile IoT sensors that are difficult to be supplied power from the connecting cables, it has received a substantial interest for researchers \cite{rana2018internet, dai2018selective, xiong2018tdma, zhang2019energy}.

One of the practical research related to RF-WPT is the Cota technology of Ossia company\cite{Ossia}. Cota technology used a beacon signal to detect energy-absorbing objects such as humans and utilized beamforming technology for transmitting the power wirelessly for a specific object while avoiding energy-absorbing objects. This company succeeded in transmitting 1 W (Watt) of Radio Frequency (RF) energy at 30 feet using 2.4GHz or 5.8GHz ISM band through Cota technology. The other company, which is called Energous, introduced a technology called WattUp. The technology uses a Wi-Fi frequency band from 5.7 GHz to 5.8 GHz to transmit 4 Watt to twelve devices at 20 feet away from the power transmitter, simultaneously.\cite{Energous}.
In addition, various RF-WPT research has been also conducted in academia. Two representative research topics are as follows: simultaneous wireless information and power transfer (SWIPT) network and the wireless powered communication network (WPCN).  

In SWIPT, the power transmitter can provides both power and information simultaneously to multiple receivers by combing a WPT system with a wireless communication system. There have been various studies related to SWIPT, such as separated SWIPT \cite{yin2016resource, feng2016distributed, liu2014secrecy,  lee2014collaborative, zhang2013mimo}, co-located SWIPT with power splitting (PS) mechanism \cite{ xu2019outage, zhou2014training, ng2013wireless, tuan2017optimal, Shi2014, zhang2015secure, zhang2013mimo, zong2015optimal}, and co-located SWIPT with time splitting (TS) mechanism \cite{xu2019outage, lee2018joint, zhang2013mimo, lee2014collaborative}.

The other application technology of WPT, where the receivers are first powered by WPT and perform their information signal transmission using the harvested energy, is referred to as WPCN. 
According to the antenna technology in WPCN, a handful of research work has been reported such as single-input single-output (SISO) WPCN \cite{ju2013throughput, chen2019resource, zhai2018accumulate}, multiple-input single-output (MISO) WPCN \cite{Xu2016, sun2014joint, Liu2014, liu2019secrecy, rezaei2019secrecy}, multiple-input multiple-output (MIMO) WPCN \cite{chen2013energy}. 

These WPT technologies provide user-friendliness, convenience, and flexibility to IoT applications because there is no need to replace the battery of IoT sensors \cite{lu2015}. 
Nevertheless, since WPT technology utilizes RF signal to transmit power, how to deal with the signal power attenuation, heavily dependent on the distance from the wireless power transmitter,  has been a severe challenge. 
For example, if IoT sensors are distributed with different distances from the wireless power transmitter, there is a huge difference in received RF power strength between IoT sensors and that induces IoT sensors to have different lifetime and available energy in the WPT system. 
More specifically, the IoT sensors, in the near of the wireless power transmitter, can harvest significantly larger energy than far away IoT sensors. 
Since IoT sensors with less available energy achieve a smaller throughput when they send their data in the uplink, H. Ju $et$ $al.$ \cite{ju2013throughput} tackled this "doubly near-far" problem as the common user throughput maximization problem and L. Liu $et$ $al.$ \cite{Liu2014} solved the problem of maximizing the minimum throughput among all users by jointly designing the DL-UL time allocation, the DL energy beamforming, and the UL transmit power allocation. 
Furthermore, since the IoT network lifetime depends on the least amount of energy of all IoT sensors, the energy fairness problems among IoT sensors should be presented to prolong the IoT network lifetime \cite{li2009network, mishra2018energy}. 
P. V. Tuan and I. Koo \cite{tuan2017optimal} solved the max-min fair harvested energy problem under the transmission power and required signal-to-interference-plus-noise-ration (SINR) constraints by combining the semidefinite relaxation (SDR) with the bisection search method. 

These works only considered the linear energy harvesting (EH) model, however, different from most existing works, the linear EH model is too ideal and a more practical nonlinear EH model should be considered \cite{valenta2014, Bosh2018}. E. Boshkovska $et$ $al.$ \cite{Bosh2015} proposed a practical nonlinear EH model based on logistic function and studied the maximization of the total harvested power at energy harvesting receivers. X. Liu $et$ $al.$ \cite{liu2019secrecy} solved the secrecy throughput maximization problem with perfect channel information state (CSI) and imperfect CSI considering nonlinear EH model in WPCNs. 
Therefore, the resource allocation algorithm, which takes into account both the energy fairness issue among IoT sensors and nonlinear EH characteristics, has been received substantial interest in researchers, recently.

This paper addresses the above issues by considering the problem of maximizing the minimum received energy of IoT sensors, which takes into account the simple logarithmic nonlinear EH model. 
In addition, we consider the IoT sensor's mobility by the random walk model and present the performance of our proposed algorithm that varies with the IoT sensor's mobility in the numerical results.

\subsection{Contribution}
In this paper, we consider a MISO-WPT system, where a single power transmitter (PT) with multiple antennas simultaneously transmits power to a number of single-antenna energy poverty IoT sensors (EPIS).  
We focus on energy fairness issues through the energy beamforming in orthogonal frequency bands and transmission power allocation under the nonlinear EH model. 
In addition, when considering the mobility of EPISs by the random walk model, the numerical results present the performance of our proposed algorithms.
The main contribution of this paper is summarized as follows: 

\begin{itemize}
\item To handle the energy fairness issue among the EPISs, the energy beamforming in orthogonal frequency bands is addressed. The EPISs located near the power transmitter receives a substantial amount of power due to their low power-signal attenuation, but relatively little power received for far away EPISs. Therefore, this paper considers the orthogonal frequency band allocation and the energy beamforming to EPISs, which received less energy.

\item We consider the practical nonlinear EH model for the reduction of misleading optimization solutions. The conventional linear EH model was adopted in most WPT works, but the recent research found that it may be an ideal concept. Different from \cite{Bosh2015} which proposed a nonlinear EH model based on logistic function, this paper considers logarithmic nonlinear EH model with operating limits, resulting in a simpler optimization problem.  Also, it allows for more accurate received power modeling. 
\item The problem of the total received power maximization (TRPM) is studied under the constraints of the limited transmission power on the orthogonal bands and the transmission power budget of PT. The closed-form solution is obtained by using the KKT optimality condition and modified water filling algorithm \cite{Park2016}. 

\item The common-received power maximization (CRPM) problem is also presented under the same condition of TRPM and solved by applying the iterative bisection search method. To apply the bisection search method to the problem, this paper proposes a method of specifying the scope of the solution for the objective function which is the form of the sum of monotonous functions.

\item Assuming the mobility of EPISs as the one-dimensional random walk model, the performance of the proposed resource allocation algorithms is demonstrated in the numeral results.

\end{itemize}

This paper is composed as follows. Section \ref{system model} discusses the system model about the WPT mechanism, which this paper proposes. Section \ref{problem} formulates the total received power maximization and common received power maximization problems and derives the optimal solutions. The numerical results of the proposed algorithms are provided in Section \ref{numerical result}. Finally, the conclusions are presented in Section \ref{conclusion}.

$Notation$  : The following notation is used in this paper. 
The uppercase bold $\boldsymbol{A}$ represents a matrix, the lowercase bold $\boldsymbol{a}$ represents a vector, the uppercase $A$ for set, and the lowercase $a$ for scalar.
For a square matrix $\boldsymbol{S}$, tr($\boldsymbol{S}$) denotes the trace of $\boldsymbol{S}$, $\boldsymbol{S}^H $ means Hermitian which denotes conjugate transpose of $\boldsymbol{S}$, 
$\lambda(\boldsymbol{S})$ denotes eigenvalues of $\boldsymbol{S}$, and $v_{\lambda}$ means eigenvector of eigenvalue of $\lambda$. The $n(A)$ denotes the cardinality of a set $A$ and 
$\|\boldsymbol{a}\|$ denotes the magnitude of a vector $\boldsymbol{a}.$

		\begin{figure}[t]
		\begin{center}
			\includegraphics[scale=0.41]{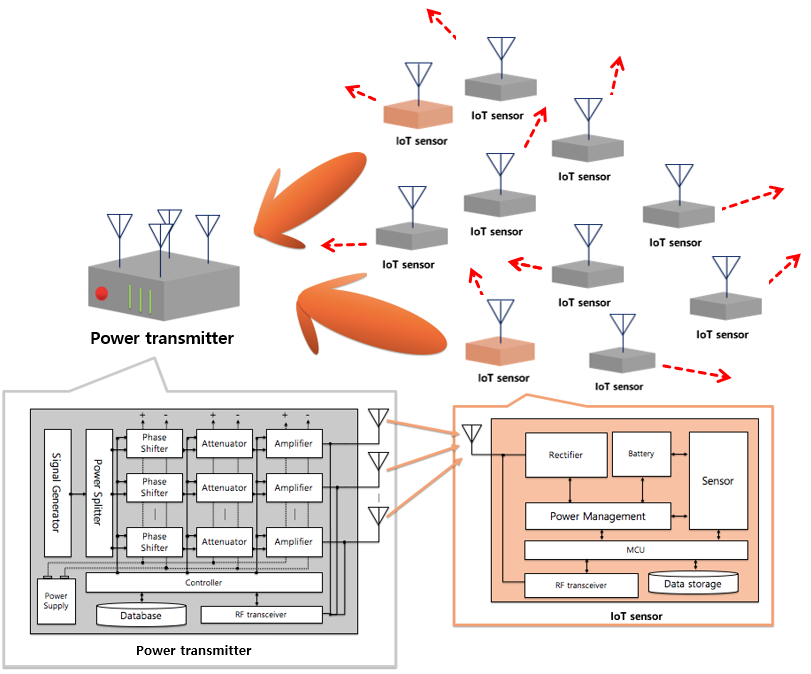}
			\caption{MISO-WPT system consisted of one power transmitter with multiple-antennas and multiple energy poverty IoT sensors (EPIS) with single antenna} \label{fig:systemmodel}
		\end{center}
	\end{figure}

	\section{System Model} \label{system model}

As shown in Fig.~\ref{fig:systemmodel}, this paper considers multi-input single-output (MISO) WPT system, consisted of one power transmitter (PT) with $n_t$ antennas and $m$ energy poverty IoT sensors (EPIS) with one antenna. The set of indices of orthogonal frequency bands are denoted by $\Gamma = \{1,2,\dots, n_C\}$. Assuming perfect channel state information (CSI), PT transmits the power to the $n_C$ targeted EPISs simultaneously. 
We suppose that all channels experience quasi-static flat fading (i.e., channel power gain on each channel stay constant within the power transmission time) and without loss of generality, the power transmission time is set to be one. Therefore, energy and power would have the same meaning.

In this paper, a plurality of EPISs requires power transmission and PT decides which EPISs should be chosen for orthogonal bands assignment and how to allocate the amount of transmission power on each orthogonal band. We design the energy beamforming in each orthogonal bands for enhancing the power transfer efficiency and consider the practical nonlinear EH model to reduce resource allocation mismatch.

	\subsection{Sensors Selection rule based on Energy Poverty (SSEP) }
	If the number of the orthogonal bands $n_C$ is larger than $m$, which is the number of EPISs, every EPISs could receive power by PT. However, if $m$ becomes larger than $n_C$, it is impossible to transmit power to all of EPISs simultaneously. Currently, since massive IoT sensors have been distributed in various IoT application domains, the number of IoT devices are generally larger than the number of the orthogonal bands. Thus, the orthogonal bands' assignment rule for EPISs is required. This paper considers IoT sensors selection rule on each orthogonal band based on energy poverty (SSEP), giving the priority to EPISs with less received energy, which allocates orthogonal band first and transmitting optimized power. In consequence, SSEP arranges EPISs according to the total received energy with ascending order and then group first $n_C$ EPISs. The EPIS's index set in this group is expressed by $\Gamma_P$. The whole procedure is expressed in Algorithm~\ref{alg:SSEP}.
	
	\begin{algorithm}[t]
		\caption{Sensor Selection rule based on Energy Poverty(SSEP)}
		\label{alg:SSEP}
		\begin{algorithmic}[1]
			\STATE \textbf{Initilization} The total received energy $U_k$ of $\text{EPIS}_k$, $\forall k \in \Gamma$, 
			and the number of orthogonal bands $n_C$.  
			\STATE Arrange EPISs according to total received energy with ascending order.
			\STATE Selects first $n_C$ EPISs. 
			\RETURN The information about selected EPISs such as total received energy, channel status, etc.
		\end{algorithmic}
	\end{algorithm}

	\subsection{Energy beamforming}
	Once EPISs are assigned with orthogonal bands through SSEP, PT considers how to transmit the RF signals for the EPISs selected. In order to enhance power transfer efficiency, we consider the energy beamforming, where power is optimally transmitted to a specific EPIS by multiplying an appropriate weight vector to the power symbol \cite{Zeng2017}.
	At first, the baseband transmitted signal $s_k$ for $\text{EPIS}_k$ is defined as
	\begin{align}  \quad\qquad\qquad \boldsymbol{s_k} = \boldsymbol{w_k}s_0,    \qquad  \quad \qquad \forall k \in \Gamma_P,
	\end{align}
	where $\boldsymbol{w_k} \!\in\! \mathbb{C}^{n_t\times1}  $ represents the beamforming weight vector for $\text{EPIS}_k$ and $s_0$ is the power symbol. It is assumed that $s_0$ is an independent and identically distributed(i.i.d.) random variable with zero mean and unit variance. 
	
	Then, the transmission power of the PT for $\text{EPIS}_k$ can be expressed as $p^{t}_{k}=E[\|\boldsymbol{s_k}\|^2]=\|\boldsymbol{w_k}\|^2 = $tr$(\boldsymbol{w_kw_k}^H) $.
	Assuming that all channels follow independent quasi-static flat fading, the channel gain vector $\boldsymbol{h_k}\!\in\! \mathbb{C}^{n_t\times1} $ is constant within power transmission time.
	The received baseband signal $\boldsymbol{y_k}$ is expressed as
	\begin{align}\qquad\qquad \boldsymbol{y_k} = \boldsymbol{h_k}^T\boldsymbol{s_k} + \epsilon_k,  \quad \qquad \forall k \in \Gamma_P, \end{align}
	where $ \epsilon_k \sim N(0,\sigma_k^2)$ is additive noise. This paper ignored the noise power in the $\text{EPIS}_k$, since it is actually a negligible amount in the EPISs \cite{Zeng2017}, \cite{wang2015}. 
	Therefore, the amount of power transferred is expressed as
	\begin{align}\qquad \qquad p^{r}_{k}= \text{tr}(\boldsymbol{G_kW_k}), \qquad\qquad \forall k \in \Gamma_P,  \end{align}
	where $\boldsymbol{G_k=h_{k}h_k}^H \!\in\! \mathbb{C}^{n_t \times n_t}$ is a channel gain matrix for $\text{EPIS}_k$ and $\boldsymbol{W_k = w_kw_k}^H \!\in\! \mathbb{C}^{n_t \times n_t}$ is a beamforming weight matrix. In \cite{Zeng2017}, the optimal energy beamforming vector is obtained as
	\begin{align} \quad \qquad  \qquad \boldsymbol{w^*_k} = \boldsymbol{v}_{\max}(\boldsymbol{h_kh_k}^H), \qquad\quad \forall k \in \Gamma_P,\quad\end{align}
	where $\boldsymbol{v_{max}}$ denotes the eigenvector corresponding to the maximum eigenvalue. Since the rank of channel gain matrix $G_k$ is one, there exists only one eigenvalue. Therefore, The received power using energy beamforming is
	\begin{align}\quad\qquad\qquad p^{r}_{k}= \lambda_kp^{t}_{k},  \qquad\qquad\qquad \forall k \in \Gamma_P, \end{align}
	where $\lambda_k = \lambda(\boldsymbol{h_kh_k}^H) $ denotes the eigenvalue of $ \boldsymbol{h_kh_k}^H $.
	Since the PT can eigen-decompose $\boldsymbol{G_k}$ to obtain eigenvalue and eigenvector, PT sets an beamforming weight vector as an eigenvector that corresponds to an eigenvalue of $\boldsymbol{G_k}$. 
	
\begin{figure}[t]
	\begin{center}
	\subfloat[]{
		\includegraphics[width=1\linewidth]{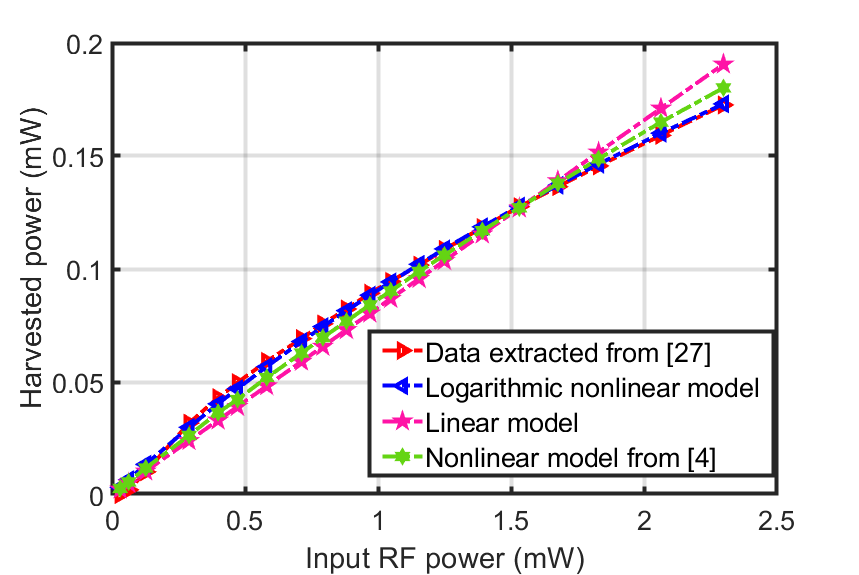}
	}
	\hskip2em
	\centering
	\subfloat[]{
		\includegraphics[width=1\linewidth]{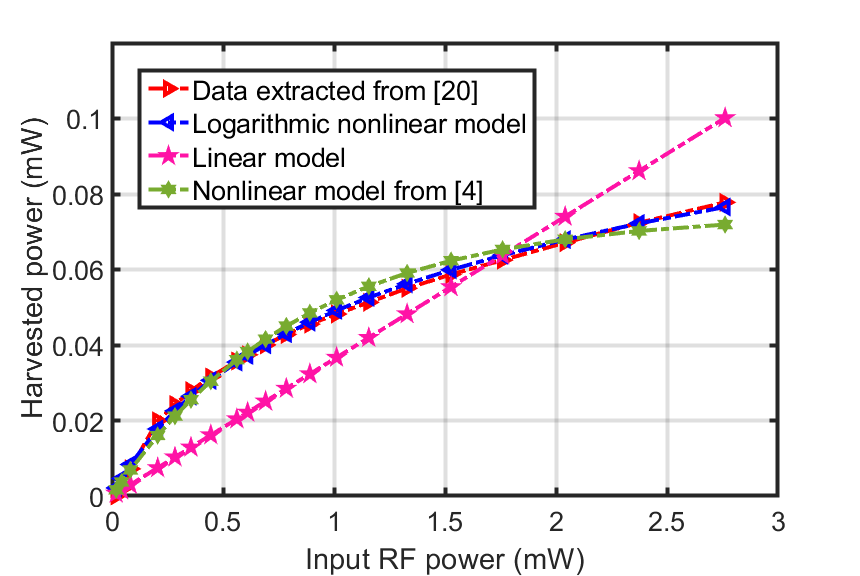}
	}
	
	\caption{(a) A comparison between linear \& nonlinear EH model  and raw data extracted from \cite{Umeda2006} and (b) a comparison between linear \& nonlinear EH models and raw data extracted from \cite{Papotto2011}. The parameters $a_k, b_k$ are calculated by minimizing the sum of squared error.} 
	\label{fig:rectifier} 
	\end{center}
\end{figure}  
	 
\begin{table}[b]
\centering	
\caption{A RMSE value comparison between the proposed logarithmic model, linear, and nonlinear model from \cite{Bosh2015} }
	
	\begin{tabular}{|
			>{\columncolor[HTML]{C0C0C0}}c |c|c|}
		\hline
		Rectifier model  & \cellcolor[HTML]{C0C0C0}\begin{tabular}[c]{@{}c@{}}RMSE value \\ from data \cite{Umeda2006} \end{tabular} & \cellcolor[HTML]{C0C0C0}\begin{tabular}[c]{@{}c@{}}RMSE value \\ from data \cite{Papotto2011} \end{tabular} \\ \hline
		Linear model                               & 0.0067                                                                                      & 0.0098                                                                                      \\ \hline
		Nonlinear model from \cite{Bosh2015}      & 0.0042                                                                                      & 0.0027                                                                                      \\ \hline
		Logarithmic nonlinear model                & 0.0027                                                                                      & 0.0020                                                                                      \\ \hline
	\end{tabular}
	\label{Tab:tabrectifier}
\end{table}

	\subsection{Nonlinear EH model } 
	When the RF signal is transmitted, it should be converted into DC power that can be used at EPISs. At this time, EPISs need the rectifier that converts the AC power into the DC power. This paper considers nonlinear rectifiers model operating in the 900 MHz frequency band and proposes a practical nonlinear EH model as
	\begin{align} \qquad r(x) = a_k\log(1+b_kx), \qquad \text{s.t.} \quad  0 \le x \le c_k,
	\end{align}
	where $a_k,b_k$, and $c_k$ are parameters that vary depending on the detailed circuit characteristics such as capacitor and diode. The $a_k, b_k$ are calculated by minimizing the sum of squared error between the data extracted using Engauge Digitizer \cite{mitchell2017engauge} and the rectifier model and $c_k$, which is rectifier's operation limit, can be easily found by checking an extracted data.
	Fig.~\ref{fig:rectifier} shows how well matches between the rectifier models and the data extracted from \cite{Umeda2006}, \cite{Papotto2011}. 
Table~\ref{Tab:tabrectifier} illustrates that the logarithmic nonlinear model best matches from the data extracted \cite{Umeda2006}, \cite{Papotto2011} in RMSE value.

\begin{algorithm}[t]
 \caption{Total Received Power Maximization (TRPM)}
\label{alg:TRPM}
 \begin{algorithmic}[1]
  \STATE \textbf{Initialization} Arrange the set of EPISs $\Gamma_P$ by the value of $\frac{1}{a_ib_i\lambda_i}$ in ascending order,\\
   the parameter indexes $j=1, k=2$, \\ 
   the waterfilling width $C_w = a_1$,  \\ 
   the remaining energy $E_r = E_c$, \\ 
   the energy height $h = \frac1{a_1b_1\lambda_1}$, \\ 
   the allocated power vector of EPISs $\boldsymbol{p^t} = 0$, \\ 
   $ a_{n_C+1}, b_{n_C+1} = 1$, \quad and \quad  $\lambda_{n_C+1} =\infty$.  \\ 
\WHILE {$E_r>0$}
 \IF { $ \bigg(    \frac{\min\left(\frac{c_k}{\lambda_k},P_c\right)}{a_j} > \frac1{a_kb_k\lambda_k} \bigg)$ \\ 
 	$ \qquad \&  \bigg(C_w\left(\frac1{a_kb_k\lambda_k}-h\right) < E_r \bigg) $ }
\STATE \text{Update} $E_r = E_r - C_w\left(\frac1{a_kb_k\lambda_k}-h\right),$ \\
$C_w = C_w + a_k, \quad h= \frac1{a_kb_k\lambda_k}, \quad $ \text{and} \quad $k = k + 1. $ 
\STATE Rearrage the set of EPISs \{j,j+1,\dots,k\} $\in \Gamma_P$ by the value of $\frac{\min\left(\frac{c_k}{\lambda_k},P_c\right)}{a_j}$ in ascending order.  \\ 
\ELSIF { $ \bigg(  \frac{\min\left(\frac{c_k}{\lambda_k},P_c\right)}{a_j}\le\frac1{a_kb_k\lambda_k} \bigg)$ \\ \qquad  \&  
  $ \bigg( C_w\left( \frac{\min\left(\frac{c_k}{\lambda_k},P_c\right)}{a_j}    -h\right) < E_r \bigg)     $ } 
\STATE \text{Update} $E_r = E_r - C_w\left(    \frac{\min\left(\frac{c_k}{\lambda_k},P_c\right)}{a_j}-h   \right),$ \\ $C_w = C_w - a_j, \quad h = \frac{ c_k }{\lambda_ka_j}, \quad$ \text{and} \quad $ j = j+1.$  
\STATE \text{Set}  $p^{t}_{j} = \frac{\min\left(\frac{c_k}{\lambda_k},P_c\right)}{a_j}.$  
\ELSE
\STATE \text{Set} $h = h + \frac{E_r}{C_w} \quad$ \text{and} $\quad E_r = 0.$  
\FOR {$i = j:k$} 
\STATE \text{Set} $p^{t}_{i}=a_i \left(h - \frac1{a_ib_i\lambda_i} \right).$ 
\ENDFOR
\ENDIF
\ENDWHILE
 \RETURN Set of allocated transmission power of EPISs $\boldsymbol{p^t}$ , after reaggraged by the original order.
 \end{algorithmic}
 \end{algorithm}

	\section{Problem formulation and solution} \label{problem}
	\subsection{Total Received Power Maximization (TRPM)}
	Considering the energy beamforming and the nonlinear EH model, PT should decide how much power needs to be transmitted for each EPISs. This part focus on maximizing the total received power of all $\text{EPIS}_k$ $\in \Gamma_P$. The following total received power maximization (TRPM) problem can be mathematically formulated as

\begin{algorithm}[h]
	\caption{Bisection method for $\alpha$}
	\label{alg:Bisection method}
	\begin{algorithmic}[1]
		\STATE \textbf{Initialization} the rectifier parameters $a_k, b_k$ for $k \in \Gamma_P$, \\
		the allocated eigenvalue ${\lambda_k}$ for $k \in \Gamma_P$,     \\
		the total received power $U_k$ for $k \in \Gamma_P$,     \\
		the remaining energy $E_r $, \\
		$\alpha_{min} = \min_k \left(U_k+ a_k\log(1+\frac{b_k\lambda_kE_r}{|\Gamma_P|})    \right)$, \\ 
		$\alpha_{max} =  \max_k \left(U_k+ a_k\log(1+\frac{b_k\lambda_kE_r}{|\Gamma_P|})    \right)$,\\
		$\epsilon$ = stopping criterion, \quad and \quad $\alpha = \frac{\alpha_{min} + \alpha_{max}}{2}$. \\
		\WHILE {$  \{|\sum_k \frac{1}{b_k\lambda_k}\left(\exp(\frac{\alpha-U_k}{a_k})-1\right)-E_r| \ge \epsilon \} $}
		\IF { $\sum_k \frac{1}{b_k\lambda_k}\left(\exp(\frac{\alpha-U_k}{a_k})-1\right)  > E_r   $ }
		\STATE $\alpha_{max} \leftarrow \frac{\alpha_{min}+\alpha_{max}  }{2}.$
		\ELSE
		\STATE $\alpha_{min} \leftarrow \frac{\alpha_{min} + \alpha_{max}}{2}.$  
		\ENDIF
		\STATE $\alpha \leftarrow \frac{\alpha_{min}+\alpha_{max}}{2}.$
		\ENDWHILE
		\RETURN The parameter $\alpha$ satisfying the condition $\sum_k \frac{1}{b_k\lambda_k}\left( \exp(\frac{\alpha-U_k}{a_k})-1 \right) = E_r $.
	\end{algorithmic}
\end{algorithm}

	\begin{align}
		\max_{\boldsymbol{p^{t}}}\sum_{k \in {\Gamma_P}}r(p^{t}_{k})=  \sum_{k \in \Gamma_P} a_k \text{log}(1+b_k\lambda_kp^{t}_{k}) \qquad \\
		\text{s.t.} \qquad  0 \le p^{t}_{k} \le  \min\left(\frac{c_k}{\lambda_k},P_c\right), \qquad  \forall k \in \Gamma_P, \\ 
		\sum_{k \in \Gamma_P}p^{t}_{k}\le E_c, \qquad\qquad \qquad\qquad  \label{totalsum}
	\end{align}
	where $E_c$ is the transmission power budget which PT has, $P_c$ is transmission power constraint under each orthogonal band, and $c_k$ is the rectifier's operation limit of $\text{EPIS}_k$. Solving the above optimization problem, we can obtain the closed-form solution given by the following theorem.  
	\begin{theorem}
	\label{thm:TRPM}
		The optimal power value $p^{t*}_{k}$ to be sent to $\text{EPIS}_k$ in the TRPM problem is expressed as follows
		\begin{align}
		\label{TRPMsol}
		p^{t*}_{k} = \begin{dcases}
		ha_k-\frac1{b_k\lambda_k},
		\quad \textrm{if this is $> 0$ or $ < \min(P_c, \frac{c_k}{\lambda_k})$}, \\
		\qquad \frac{c_k}{\lambda_k},\qquad   \textrm{if  the above value is $\ge \min(P_c, \frac{c_k}{\lambda_k})$},  \\
		\quad\quad   0,  \qquad \qquad \text{Otherwise,}
		\end{dcases}
		\end{align}
		where $h$ is the lagrange multipier satisfying $\sum_{k \in \Gamma_P}p^{t*}_{k} = E_r$.

	\end{theorem}
	
\begin{IEEEproof} See the Appendix \ref{appendix:KKT}.  \end{IEEEproof}

	Through the $Theorem$ \ref{thm:TRPM}, the optimal power value $p^{t*}_k$ for $\text{EPIS}_k$ is obtained. The $h$ is calculated by Alogorithm~\ref{alg:TRPM}, which is the modified waterfilling algorithm \cite{Park2016}. 
	If PT transmits power $p^{t*}_k$ to $\text{EPIS}_k,$ $\forall k \in \Gamma_P$, then total received power for all $\text{EPIS}_k,$ $\forall k \in \Gamma_P,$ is maximized.

\subsection{Common Received Power Maximization (CRPM)}
PT could utilize the TRPM algorithm to optimize the total received power of all $\text{EPIS}_k,$ $\forall k \in \Gamma_P$.  However, it can cause the energy unfairness issue among the EPISs. Through SSEP, EPISs with less received energy are given the opportunity to receive power, but in TRPM, most of the power is transmitted to the EPIS, which is the nearest PT, causing an energy imbalance. Therefore, it is required that a sophisticated resource allocation algorithm without causing an energy imbalance among EPISs.
	This part considers Common Received Power Maximization (CRPM) problem to solve this energy unfairness issue. It can be formulated as
	\begin{align}
	\max_{\boldsymbol{p^{t}},\bar{P}} \quad \bar{P} \qquad\qquad \qquad \qquad \\
	\text{s.t.} \label{pconstraint1} \quad U_k+a_k\log(1+b_k\lambda_k p^{t}_{k}) \ge \bar{P}, \quad \forall k \in \Gamma_P,\quad \\ 
	\label{pconstraint2}\qquad\quad  0 \le p^{t}_{k} \le  \min\left(\frac{c_k}{\lambda_k},P_c\right),  \quad \forall k \in \Gamma_P, \quad  \\ 
	\label{totalequal} \quad \sum_{k \in \Gamma_P}p^{t}_{k} = E_c,  \quad  \qquad\qquad  \qquad
	\end{align} 
	where $\bar{P}$ is the common received power for all EPISs. Here, the transmission power budget condition (\ref{totalequal}) has equal value to the constraint $E_c$. It is natural to consume all of the transmission power budgets at PT because it can improve the performance of common received power $\bar{P}$ of EPISs.  Since (\ref{pconstraint1}) constraint is increasing about $p^{t}_{k}$, when all these conditions have equal value, the optimal solution of CRPM problem can be obtained as follows 

	\begin{align}
	\qquad  U_k+a_k\log(1+b_k\lambda_kp^{t}_{k}) = \alpha, \qquad \forall k \in \Gamma_P. \\ 
	\textrm{Then,} \quad
	p^{t}_{k} = \frac1{b_k\lambda_k}\left(\exp(\frac{\alpha-U_k}{a_k})-1  \right). \qquad  
	\end{align}
	Since $p^{t}_{k}$ is increasing and continuous function over $\alpha$, we can find $\alpha$ by using bisection search method. The interval in which solution $\alpha$ exists is defined by following lemma. 
	\begin{lemma} 
	\label{lemma:1}
		The solution $\alpha$ satisfying the condition (\ref{totalequal}) exists in the following interval $\alpha \in $ $\left[ \alpha_{\min}, \alpha_{\max}\right] $ where 
\begin{center} $\alpha_{\min} =  \min_{k \in \Gamma_P} \left(U_k+a_k\log(1+ \frac{b_k\lambda_kE_c}{n(\Gamma_P)} )       \right), $ \\
$\alpha_{max} =  \max_{k \in \Gamma_P} \left(U_k+a_k\log(1+ \frac{b_k\lambda_kE_c}{n(\Gamma_P)} )   \right).  $   \end{center}

	\end{lemma}
	\begin{IEEEproof} See the Appendix \ref{appendix:SSSM}.  \end{IEEEproof}
	
	Since we should consider the constraints about $p^{t}_k$ such as (\ref{pconstraint1}) and (\ref{pconstraint2}), the solution of the problem is composed of the iterative bisection search algorithm. The whole procedure is summarized in Algorithm~\ref{alg:CRPM}.

\begin{algorithm}[t]
	\caption{Common Received Power Maximization (CRPM)}
	\label{alg:CRPM}
	\begin{algorithmic}[1]
		\STATE \textbf{Initialization} The number of iteration $I$= 0,\\
		the limit of Iteration $I^{\text{lim}}$, \\
		the remaining energy $E_r = E_c$, \\  
		the allocated power vector $\boldsymbol{p^t} = 0$, \\ 
		and the index set of EPISs $\Gamma_P. $ \\
		\WHILE {$\{E_r > 0\}$  and  $\{I < I^{\text{lim}}     \}  $ }
		\STATE Calculate $\alpha$ by Algorithm~\ref{alg:Bisection method} 
		\FOR {$ k \in \Gamma_P$}
		\STATE Set $p^{t}_{k} = p^{t}_{k} + \frac{1}{b_k\lambda_k}\left(\exp(\frac{\alpha-U_k}{a_k})-1   \right). $
		\ENDFOR
		\STATE Set $E_r$ = 0.
		\FOR {$ k \in \Gamma_P$}
		\IF { $p^{t}_{k} > \min\left(\frac{c_k}{\lambda_k},P_c\right)   $ }
		\STATE Set $E_r = E_r + p^{t}_{k} - \min\left(\frac{c_k}{\lambda_k},P_c\right),   $\\
		$p^{t}_{k} = \min\left(\frac{c_k}{\lambda_k},P_c\right)$, \quad and \quad $\Gamma_P \leftarrow  \Gamma_P \setminus k.$
		\ENDIF
		\ENDFOR
		\STATE $I = I+1.$ 
		\ENDWHILE
		\RETURN Allocated transmission power vector $\boldsymbol{p^t}$.
	\end{algorithmic}
\end{algorithm}

\section{Numerical Results} \label{numerical result}
\subsection{Simulation Setup}
In this section, we provide a numerical example to identify the performance of the proposed resource allocation algorithms. We consider a WPT system of 16-mobile energy poverty IoT sensors (EPIS)s with a single antenna and one-power transmitter (PT) with four-antennas. 
We assume the following a few numerical simulation settings:
the PT performed a total ten-thousand iteration of wireless power transmissions and at each time the maximum transmission power $E_c$ of the PT and the limited transmission power constraint $P_c$ in each orthogonal band is assumed to be 4W (Watt). The transmission power constraints are determined by reflecting the power regulation of the 900MHz band, which is the frequency band used for IoT communication. 
The nonlinear rectifier parameters $a_k, b_k$ of the k-th EPIS are determined randomly between two values,  
a = 0.0319, b = 3.6169 from \cite{Umeda2006} and a = 0.2411 and b = 0.4566 from \cite{Papotto2011}. 
The paired values in a and b are obtained by minimizing the mean squared error (MSE) between the nonlinear EH model and extracted data from \cite{Umeda2006} and \cite{Papotto2011}, respectively.
The $c_k$, which denotes the operation limit of the rectifier, is fixed to 3 mW for both options. 
Within a 5 to 15 meter from PT, each EPISs is uniformly distributed in the WPT system and randomly moves with fixed-length walk-steps (0.03 m or 0.2 m)  in each iteration. 
At each iteration of power transmission, we consider a one-dimensional random walk model of EPISs. It has three options to stop, move forward, and move backward based on PT, and the options have an equal probability, which is 1/3.
 The distance-dependent signal pathloss model is given by following equation. 
\begin{align*}
	L_k = L_0\bigg( \frac{d_k}{d_0}   \bigg)^{-\alpha}, \qquad \forall k \in \Gamma_P,
\end{align*}
where $L_0$ is set to be $10^{-3}$, $d_k$ denotes the distance between PT and $\text{EPIS}_k$, $\forall k \in \Gamma_P$, $d_0$ is a reference distance, which is 1 meter, and $\alpha$ is the path loss exponent set to be 3.
The channel vector $h_k$ is obtained by averaging over 1,000 randomized independent and identically distributed (i.i.d.) Rayleigh fading channel generations and the average power of $\boldsymbol{h_k}$ is normalized by $L_k$.  Assuming that all channels follow independent quasi-static flat fading, the channel gain vector $\boldsymbol{h_k} $ is constant within the power transmission time. 

In this paper, we compare the proposed SSEP with the Round-robin scheduling algorithm, where the orthogonal bands are assigned to each EPIS in equal portions and in sequential order without priority. It guarantees the fairness for the number of the power receiving opportunities among EPISs. In addition, we validate the performance of the proposed power allocation algorithms by comparing the equal power distribution (EPD) algorithm additionally, where transmission power is divided equally on each orthogonal band. 

In order to identify the performance of the proposed power allocation algorithms in the nonlinear EH model, we study the problems in the linear EH model as follows: the Linear Total Received Power Maximization (LTRPM) problem and the Linear Common Received Power Maximization (LCRPM) problem. First, the LTRPM problem is defined as

\begin{align}
\max_{\boldsymbol{p^{t}}}\quad \sum_{k \in {\Gamma}}h_k \lambda_k p^{t}_{k}\qquad\qquad \qquad\quad  \\
\text{s.t.} \quad  0 \le p^{t}_{k} \le \min\left(\frac{c_k}{\lambda_k},P_c\right), \quad  \forall k \in \Gamma_P, \\ 
\sum_{k \in \Gamma_P}p^{t}_{k} = E_c, \qquad\qquad \qquad \qquad 
\end{align}
where $h_k$ is a linear rectifier parameter derived by minimizing the mean square error between the data and linear rectifier model and $E_c$ is total transmission power constraint. Since this is a linear optimization problem, it can be solved by applying the sequential power allocation method. First, sorting the $h_k\lambda_k$ by descending order, and then allocating power $p^{t}_{k} = \min(P_c, \frac{c_k}{\lambda_1},E_r)$, $\forall k \in \Gamma_P$, in sequential order until the remaining energy $E_r$ equals to zero. It can be expressed as the Algorithm \ref{alg:LTRPM} in Appendix \ref{APPENDIX LTRPM}. 

Next, the LCRPM problem is defined as

\begin{align}
\max_{\boldsymbol{p^{t}},\bar{P}} \quad \bar{P} \qquad\qquad\qquad \qquad \qquad \\
\text{s.t}  \qquad \qquad U_k+h_k \lambda_k p^{t}_{k} \ge \bar{P},  \quad \quad  \qquad  \forall k \in \Gamma_P, \quad \\ 
\quad 0 \le p^{t}_{k} \le \min\left(\frac{c_k}{\lambda_k},P_c\right),   \quad \forall k \in \Gamma_P, \quad  \\ 
\quad \sum_{k \in \Gamma_P}p^{t}_{k} = E_c.  \quad  \quad\qquad\qquad \qquad \qquad
\end{align} 
As described above, the LCRPM problem can be solved by following the same process in the CRPM problem as follows

\begin{align}
\qquad  h_{k}\lambda_{k}p^{t}_{k} = \alpha \quad  \Longleftrightarrow \quad  p^{t}_{k} = \frac{\alpha}{h_k\lambda_k},  \qquad  \forall k \in \Gamma_P. 
\end{align}
The optimal solution can be obtained by applying the iterative bisection search algorithm, until $\sum_{k \in \Gamma_P}p^{t}_{k} = E_c$. The whole procedure is summarized as Algorithm \ref{alg:LCRPM} in Appendix \ref{APPENDIX LCRPM}. 

\begin{figure}[t]
	\centering
	\includegraphics[scale=0.25]{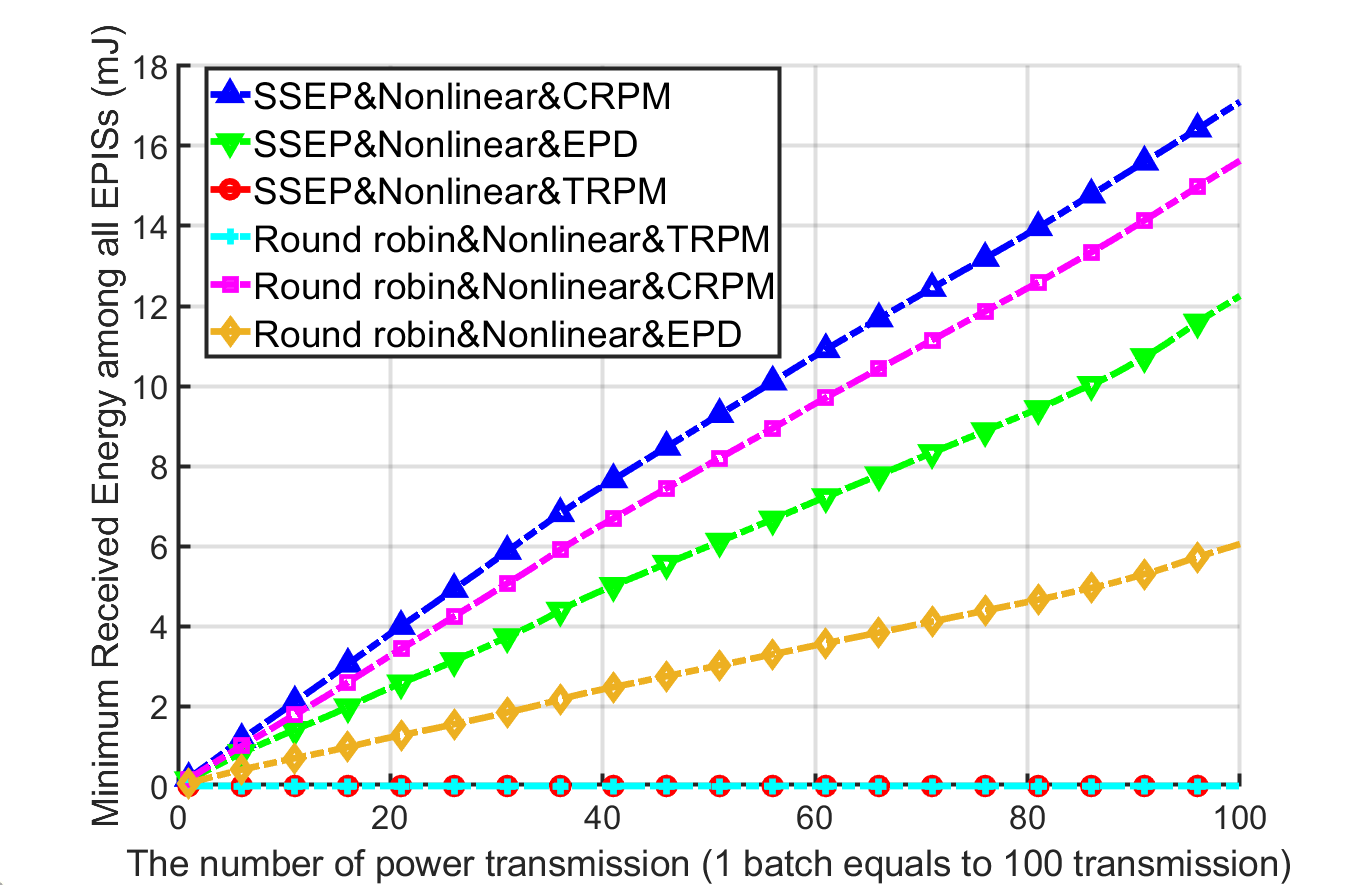}
	\caption{Minimum received energy comparison versus the number of power transmission under various power and orthogonal bands allocation methods with walk step = 0.03 m/iteration} \label{fig:LPowerComp}
\end{figure} 
\subsection{Performance Comparison with walk-step: 0.03 m/iteration}
Fig.~\ref{fig:LPowerComp} compares the performance of resource allocation schemes, in terms of the minimum received energy of all EPISs, versus the iteration of the power transmission in 0.03 m/iteration walk-step. 
It can be observed that the SSEP has always a higher minimum received energy value than the Round-robin scheduling, and CRPM also outperforms the other power allocation methods, TRPM, EPD.  
Therefore, the SSEP\&CRPM showed the best performance in the minimum received energy. The reason for this result is that CRPM \& SSEP allocates more resources, such as transmission power and the number of transmission power received, for the EPISs with poor wireless link state. In TRPM, regardless of whether SSEP or Round-robin is used, the minimum received energy value is zero, since the power transmission has not occurred to several EPISs with a poor channel condition to maximize the total received power. 
In both the CRPM and EPD, it can be seen that the minimum received energy value increases stably without significant changes.
 
 \begin{figure}[t]
 	\begin{center}
 		\includegraphics[scale=0.25]{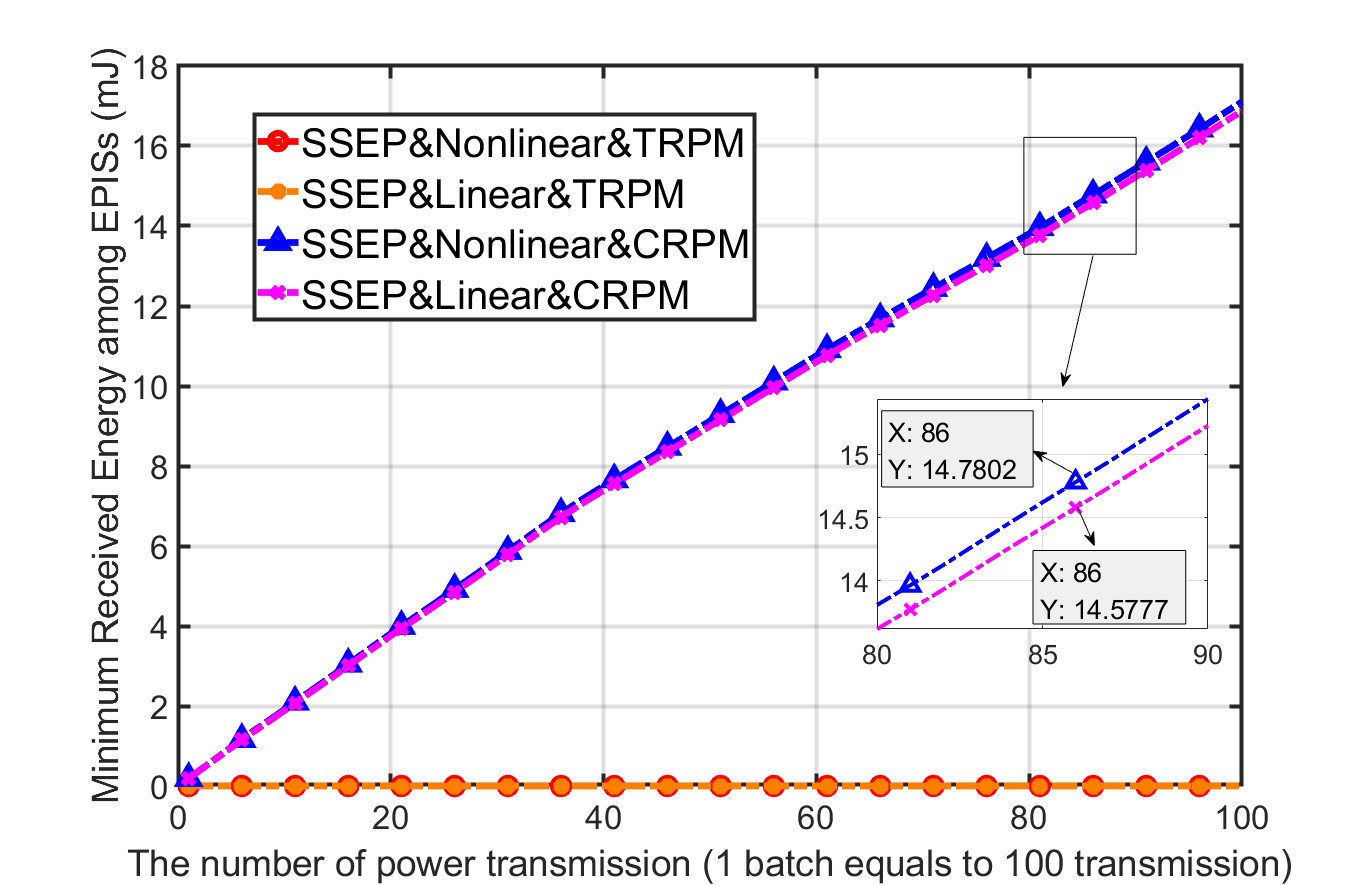}
 		\caption{Minimum received energy comparison versus the number of power transmission under linear and nonlinear CRPM \& TRPM model with walk step = 0.03 m/iteration} \label{fig:LNvsL}
 	\end{center}
 \end{figure}

In Fig.~\ref{fig:LNvsL}, we make a comparison of the performance in terms of the minimum received energy between the proposed CRPM and TRPM with considering the linear energy harvesting (EH) model and nonlinear EH model.
In the case of CRPM, it shows the nonlinear EH model achieves 1.39\% higher minimum received energy value than the linear model at the 86th batch iteration. As the number of power transmission increases, it can be seen that the gap of the minimum received energy, between the nonlinear and linear model, widens slighter than before. In TRPM, the nonlinear EH model is not comparable with linear EH model, since an EPIS that have not received power exists.


\begin{figure}[t]
	
	\centering
	\subfloat[]{
		\includegraphics[width=1\linewidth]{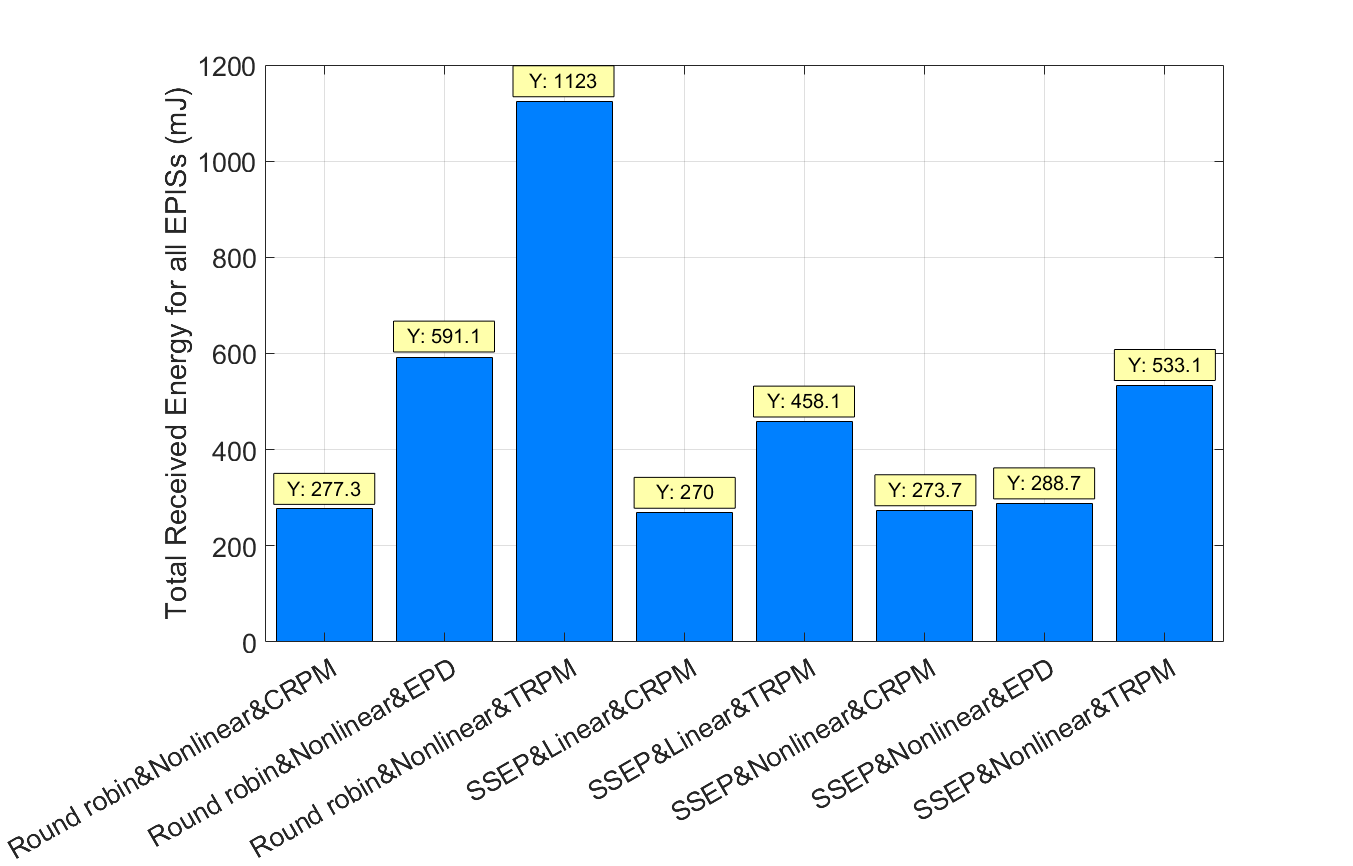}
	}
	\hskip2em
	\centering
	\subfloat[]{
		\includegraphics[width=1\linewidth]{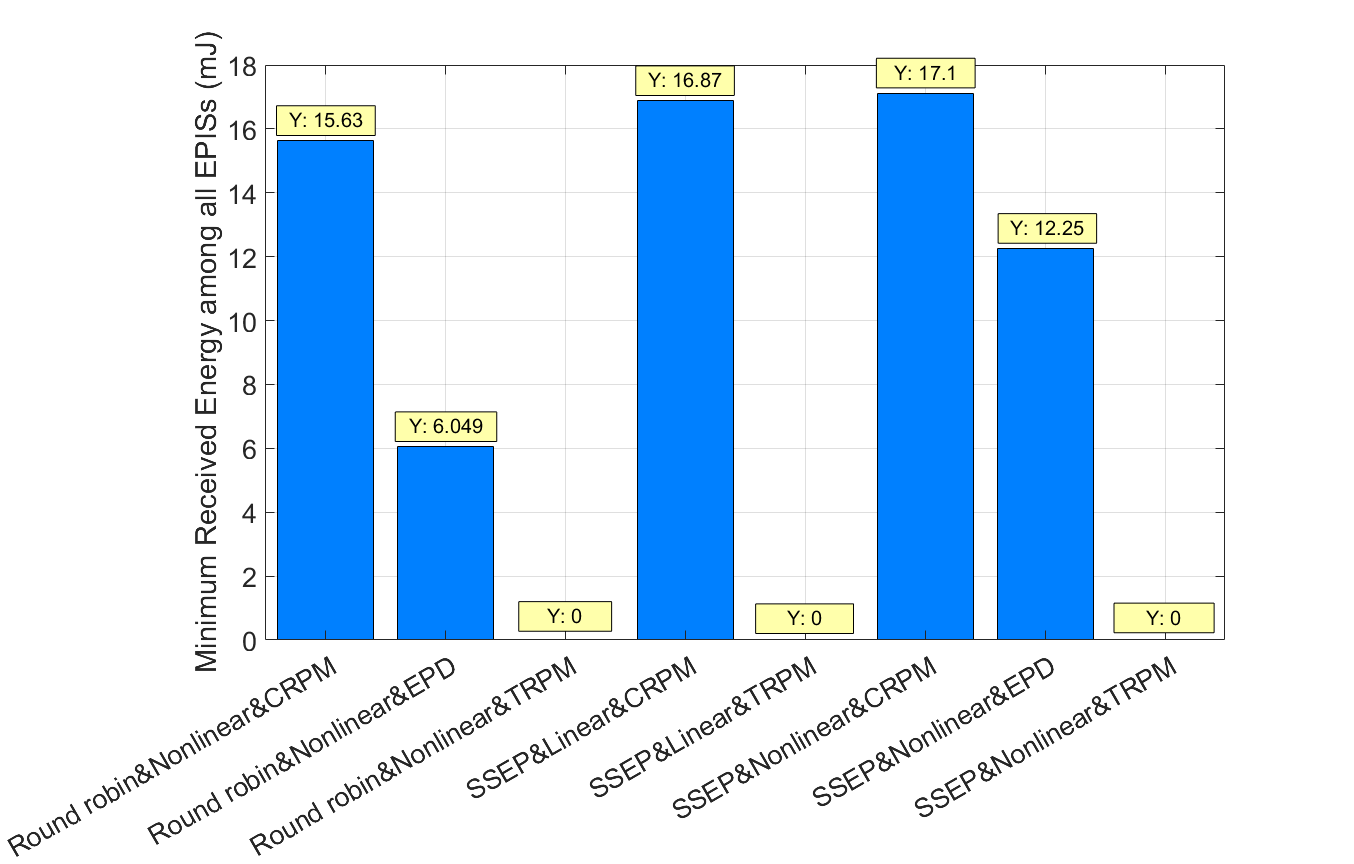}
	}
	
	\caption{(a) Total received energy comparison with walk step: 0.03m (b) Minimum received energy comparison with walk step: 0.03 m/iteration} 
	\label{fig:Ltotalandmin}
\end{figure}  

Fig.~\ref{fig:Ltotalandmin} shows the total and minimum energy received by all EPISs under different transmission power and orthogonal bands allocation schemes, for the number of 10,000 power transmission iteration.
In power allocation algorithms comparison, the TRPM shows the best performance in terms of the total received energy of all EPISs, regardless of whether SSEP or Round-robin. However, when observing the performance of minimum received energy, it can be easily found that certain EPISs should be sacrificed to maximize the total received energy of all EPISs in TRPM. In contrast to the TRPM, the total received energy for all EPISs represents the smallest value, but the best performance for the minimum received energy is detected for CRPM. In orthogonal bands allocation schemes comparison, the SSEP shows smaller performance in terms of the total received energy of all EPISs compared to the Round-robin, but better with respect to minimum received energy. In particular, the SSEP\&CRPM shows 9.4\% better performance in the minimum received energy compared to the Round robin\&CRPM, and 1.31\% less in total received energy. This trade-off result comes from that the Round-robin equally distributed the number of power transfer iteration to EPISs. It increases the probability of obtaining the energy of some EPISs that can harvest a large amount of power, but on the contrary, not elaborately takes into account EPISs, which cannot harvest large power due to the poor wireless link state.
When comparing nonlinear and linear models in the SSEP\&CRPM scheme, the minimum received energy of the nonlinear model was 1.36\% higher than the linear model and the total energy was 1.37\% higher. This result comes from the more sophisticated received power modeling achieved in the nonlinear model for each EPIS.

\begin{figure}[t]
	\centering
	\subfloat[]{
		\includegraphics[width=0.32\linewidth]{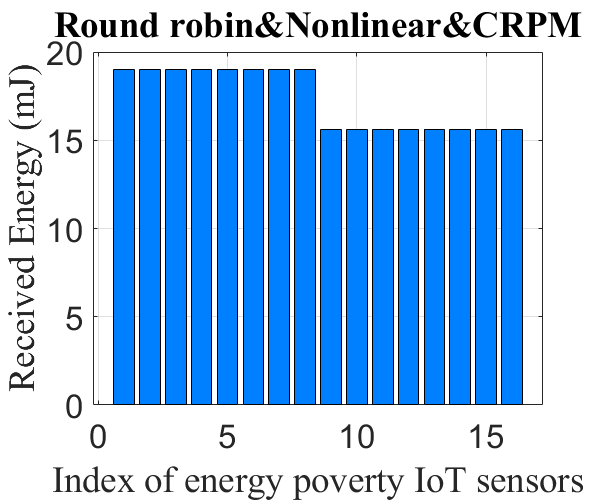}
	}
	\centering
	\subfloat[]{
		\includegraphics[width=0.32\linewidth]{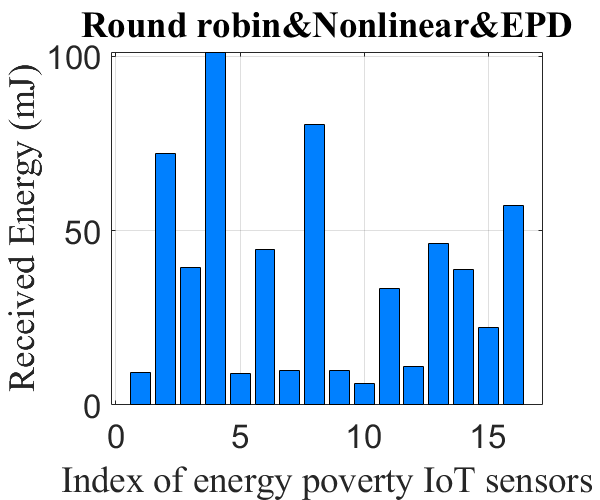}
	}
	\centering
	\subfloat[]{
		\includegraphics[width=0.32\linewidth]{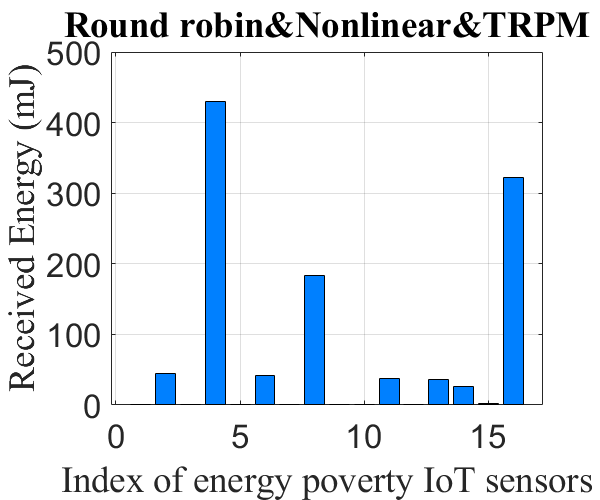}
	}
	\hskip2em
	\centering
	\subfloat[]{
		\includegraphics[width=0.32\linewidth]{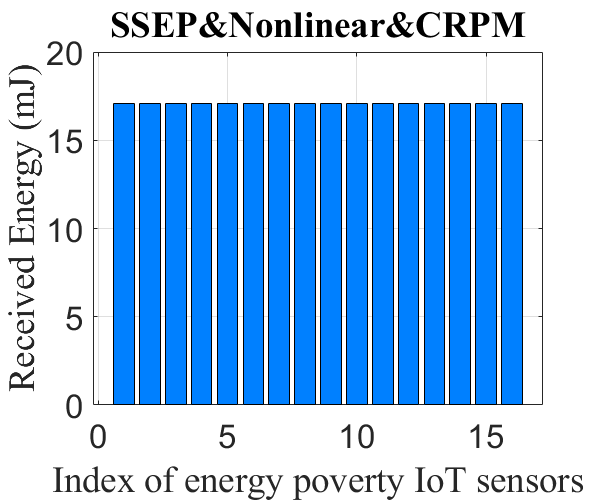}
	}
	\centering
	\subfloat[]{
		\includegraphics[width=0.32\linewidth]{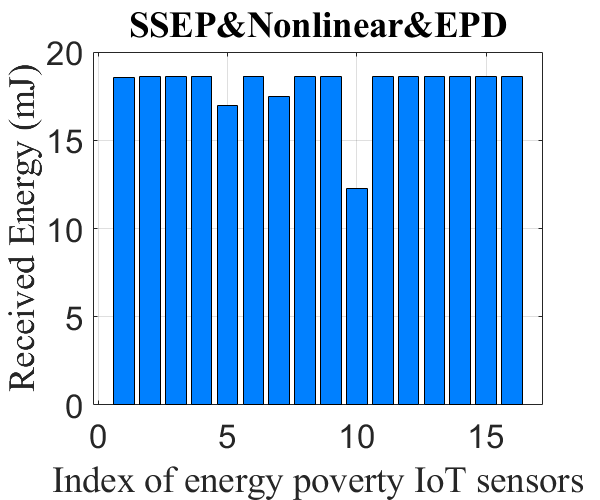}
	}
	\centering
	\subfloat[]{
		\includegraphics[width=0.32\linewidth]{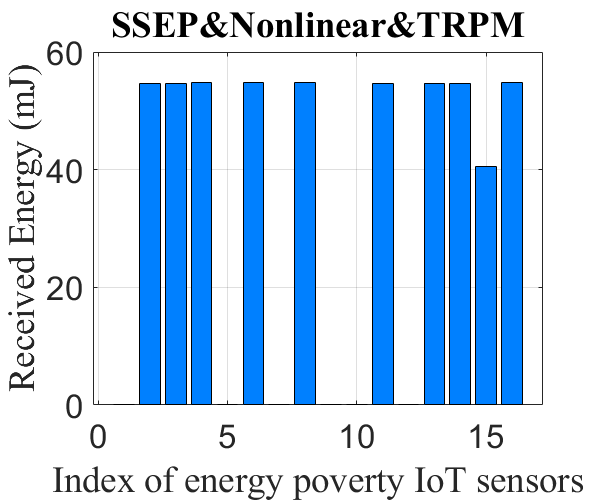}
	}
	\hskip2em
	\centering
	\subfloat[]{
		\includegraphics[width=0.32\linewidth]{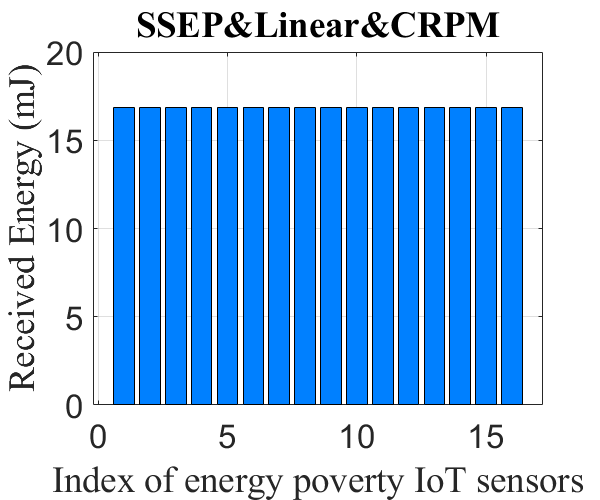}
	}
	\centering
	\subfloat[]{
		\includegraphics[width=0.32\linewidth]{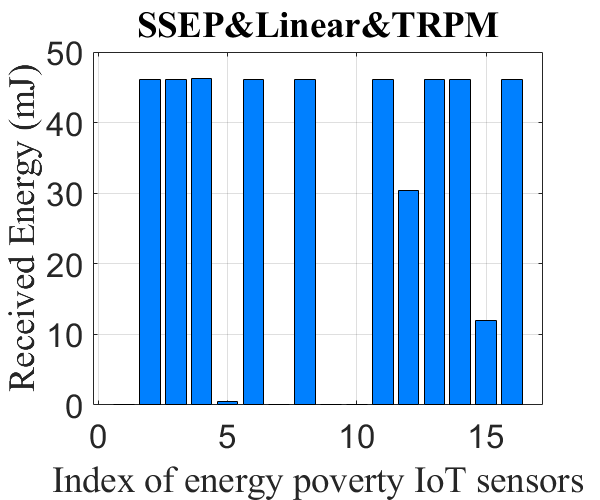}
	}
	
	\caption{Total received the energy of EPISs according to the sensor selection and power allocation methods with walk step 0.03 m/iteration\\(a) Round-robin\&Nonlinear\&CRPM \\ (b)~Round-robin\&Nonlinear\&EPD \\ (c) Round-robin\&Nonlinear\&TRPM \\(d) SSEP\&Nonlinear\&CRPM\\ (e) SSEP\&Nonlinear\&EPD\\ (f) SSEP\&Nonlinear\&TRPM \\(g) SSEP\&Linear\&CRPM \\ (h) SSEP\&Linear\&TRPM} 
	\label{fig:LowPattern}	
	
\end{figure}  

Fig.~\ref{fig:LowPattern} displays the total received energy among all EPISs with 0.03 m/iteration walk-step for the number of 10,000 power transmission iteration, under different transmission power and orthogonal bands allocation schemes. 
In Round-robin\&CRPM, the EPISs are grouped into the number of orthogonal bands and the grouped EPISs receive the same power. Thus, the different pattern of received energy which the two groups get is observed as shown in Fig.~\ref{fig:LowPattern}(a). In TRPM, since it maximizes the total received energy of all EPISs, a huge difference in total received energy between EPISs is observed. In particular, Fig.~\ref{fig:LowPattern}(f) and (h) show that the 12th EPIS cannot receive power on the nonlinear model, but it receives a certain amount of power from the linear model. This can be seen as a result of misleading optimized power allocation, due to the lack of accurate modeling in received power.
When comparing nonlinear and linear EH models in CRPM, significant performance degradation is not observed in the linear EH model.

\begin{figure}[t]
	\begin{center}
		\includegraphics[scale=0.20]{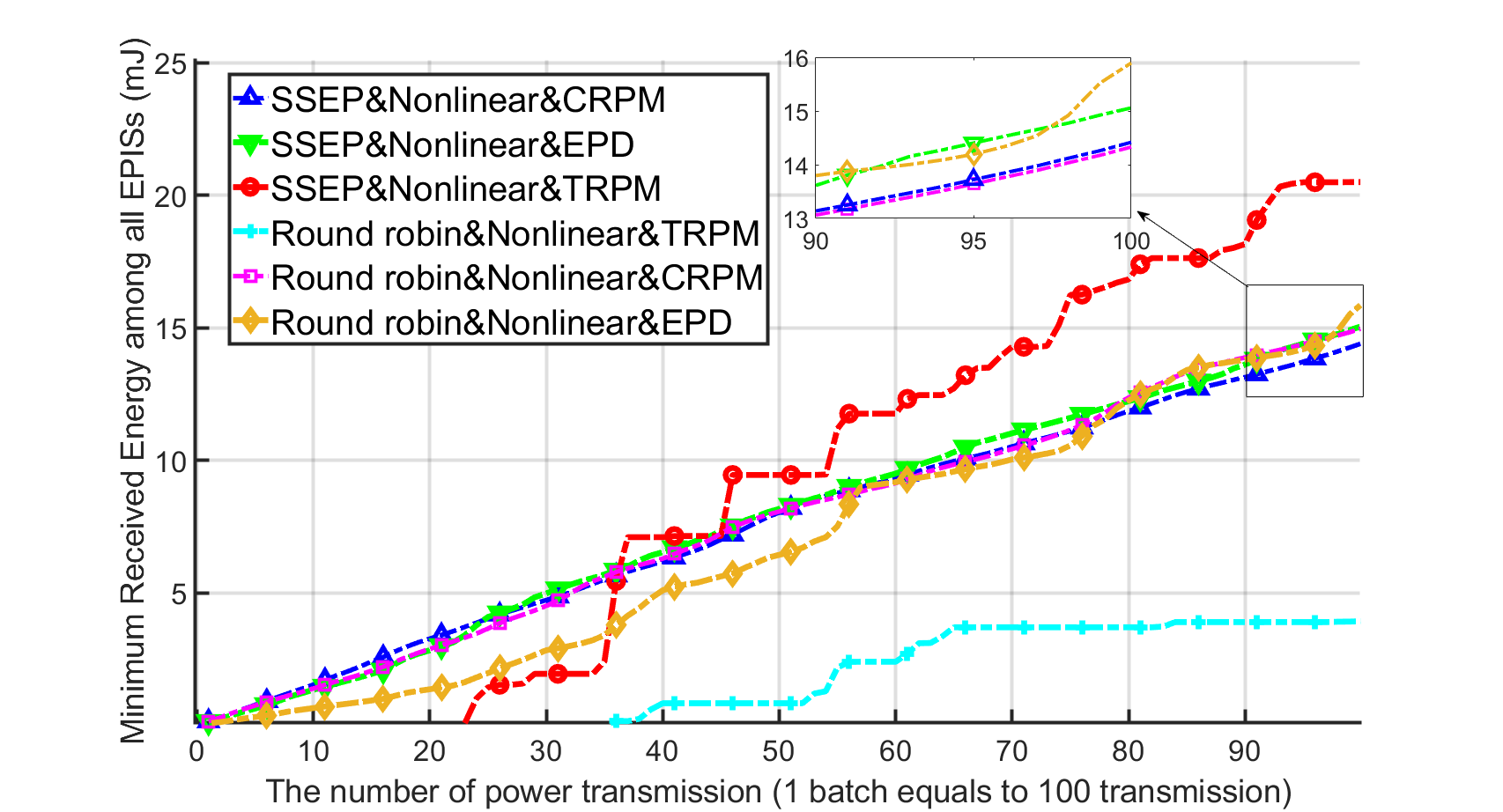}
		\caption{Minimum received energy comparison versus the number of power transmission under various power and orthogonal bands allocation methods with walk step = 0.2 m/iteration} \label{fig:HPowercomparison}
	\end{center}
\end{figure} 
\subsection{Performance Comparison with walk-step: 0.2 m/iteration}
Fig.~\ref{fig:HPowercomparison} compares the performance of resource allocation schemes, in terms of the minimum received energy of all EPISs, versus the iteration of the power transmission in 0.2 m/iteration walk-step. 
In TRPM, the minimum received energy value ascends like stepwise as the number of power transmission increases. In addition, in SSEP, it is observed that CRPM had lower minimum received energy than the other two power allocation algorithms, TRPM and EPD, unlike the result of 0.03 m walk-step. The reason for this result is that the EPISs, which received less energy due to the poor channel state, gain more improved wireless link-state by moving more dynamically than 0.03 m/iteration walk-step and it leads to enhancing the probability of obtaining a substantial power. Nevertheless, there are no significant performance differences between different power allocation algorithms, except TRPM.

\begin{figure}[t]
	\centering
		\includegraphics[scale=0.20]{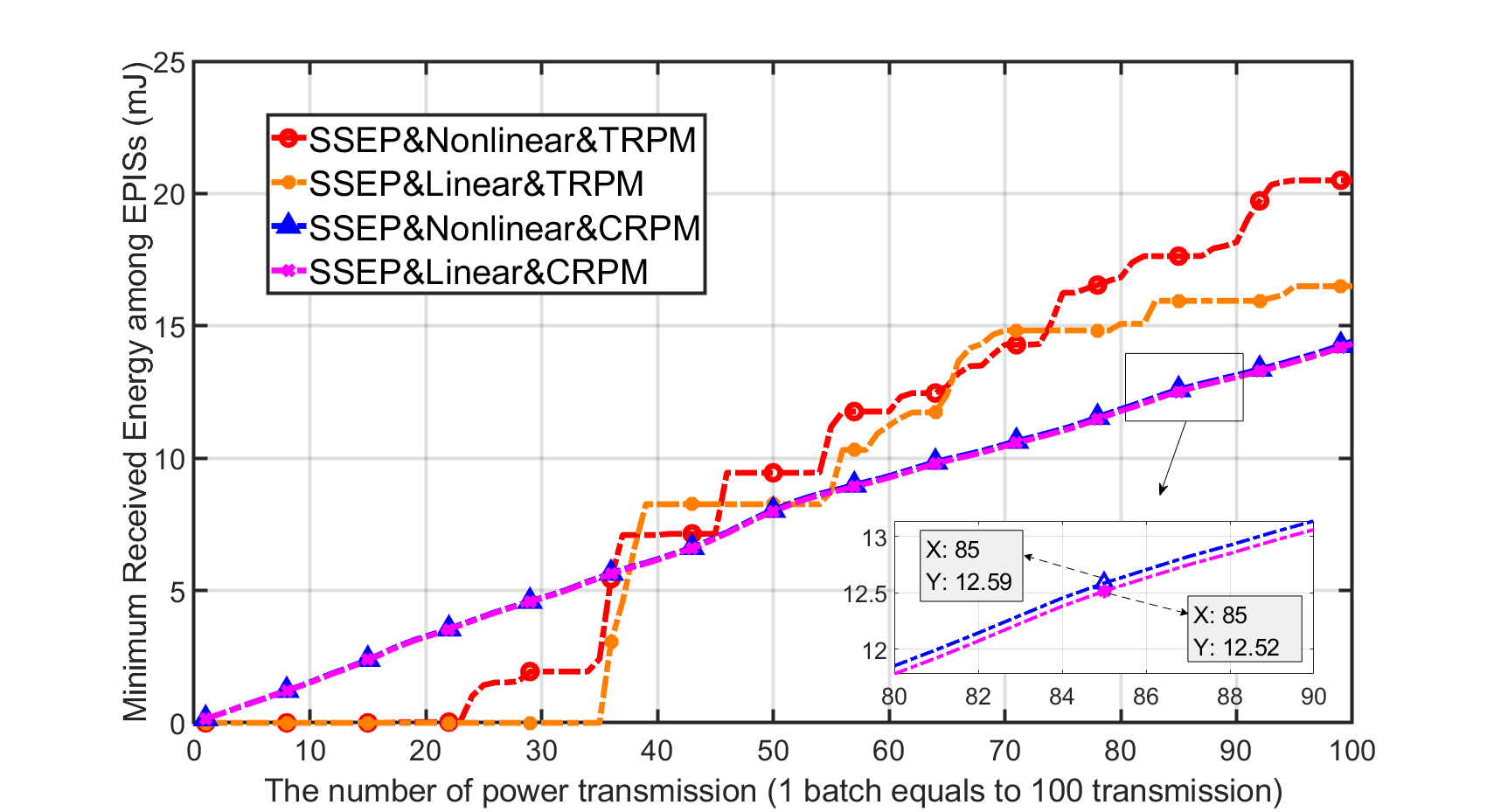}
		\caption{Minimum received energy comparison versus the number of power transmission under linear and nonlinear CRPM \& TRPM model with walk step = 0.2 m/iteration} \label{fig:HNvsL}
\end{figure} 

\begin{figure}[h]
	\centering
	\subfloat[]{
		\includegraphics[width=1\linewidth]{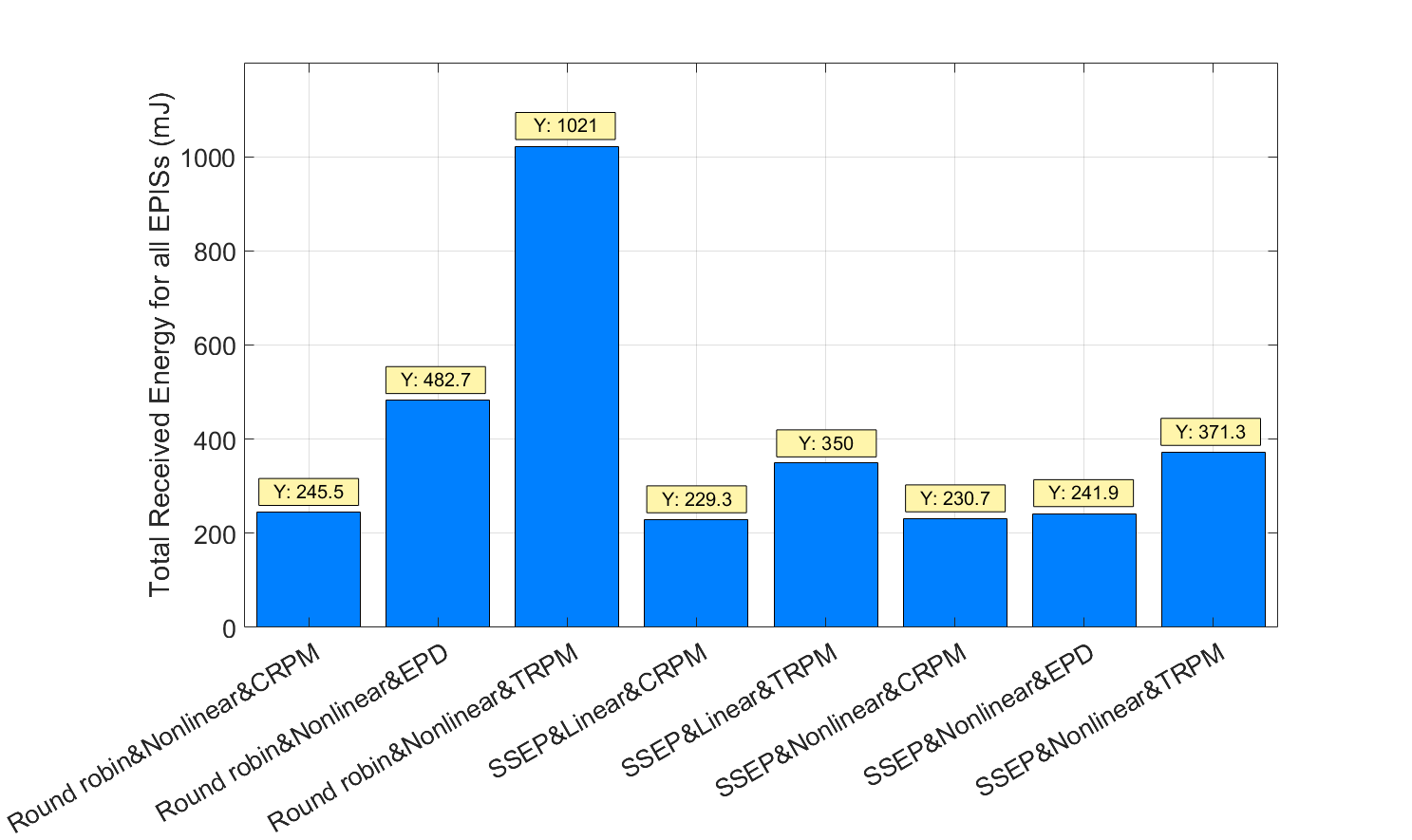}
	}
	\hskip2em
	\centering
	\subfloat[]{
		\includegraphics[width=1\linewidth]{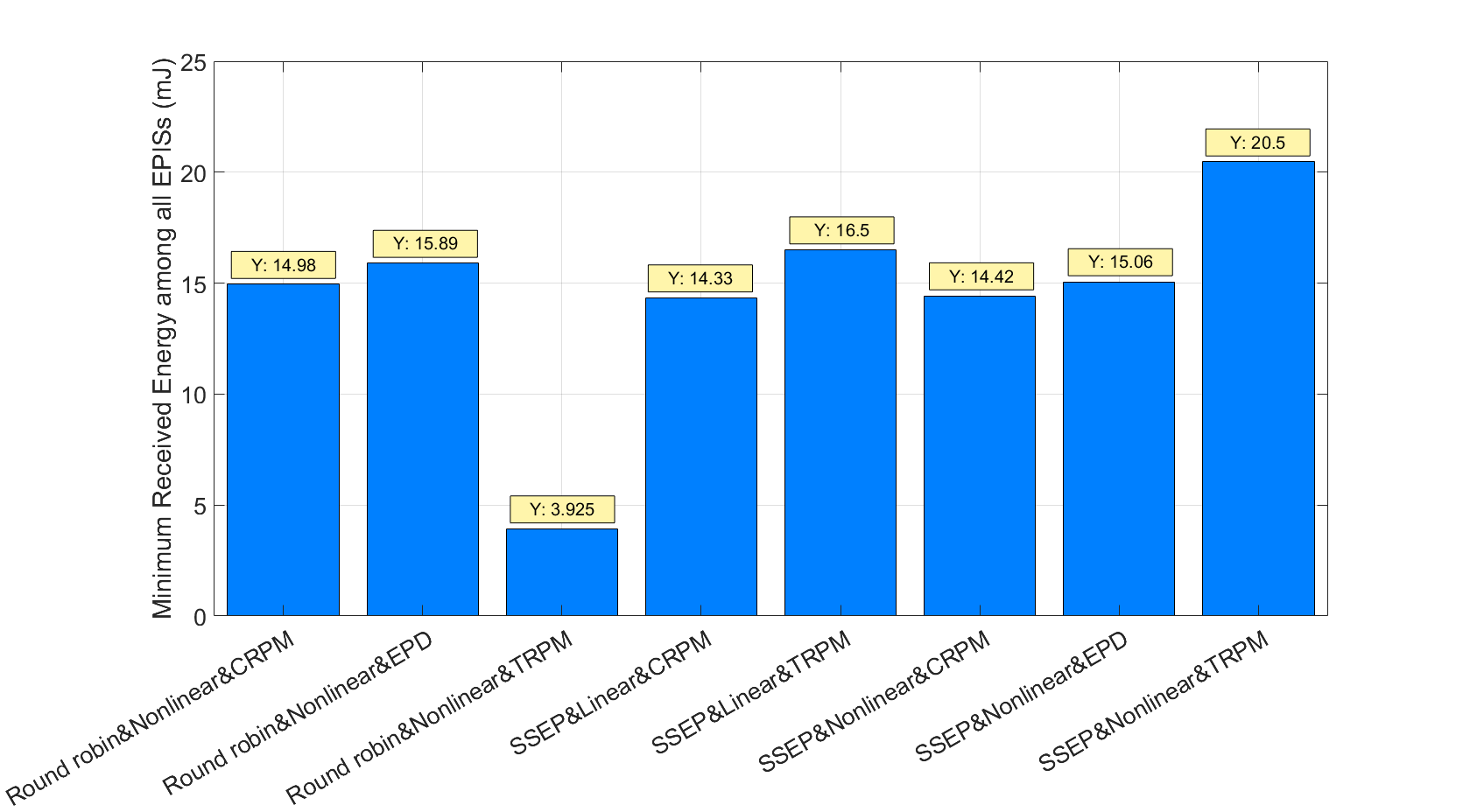}
	}
	
	\caption{(a) Total received energy comparison with 0.2 m walk step (b) Minimum received energy comparison with 0.2 m/iteration walk-step} 
	\label{fig:Hmintotal}
\end{figure}  
We further conduct the performance comparison of the minimum received energy versus the iteration number of power transmission, in the CRPM and TRPM algorithms, between the linear and nonlinear EH model, as shown in Fig.~\ref{fig:HNvsL}.
It is observed that althrough the performance of the linear EH model is close to that of the nonlinear EH model in CRPM, the nonlinear EH model always outperforms the linear EH model slightly, in terms of the minimum received energy.
When x is 85 batch iteration, the minimum received energy is about 0.6\% higher in the nonlinear EH model of CRPM.
Furthermore, in both nonlinear and linear EH models in TRPM, it can be seen that the minimum received energy value ascends like stepwise, due to the movement of EPISs. As the number of power transmission increases, the nonlinear EH model outperforms the linear model.

Fig.~\ref{fig:Hmintotal} shows the total and minimum received energy by all EPISs for the number of 10,000 power transmission iteration, under different transmission power and orthogonal bands allocation schemes, at 0.2 m/iteration walk-step. In SSEP, TRPM shows the best performance in both total and the minimum received energy among all proposed power allocation algorithms. TRPM is shown to be 42\% higher than CRPM at minimum received energy and 61\% higher at total received energy. 
Although Round robin\&TRPM is approximately 175\% higher than SSEP\&TRPM at total energy, at minimum received energy, SSEP\&TRPM showed approximately 422\% higher performance than Round robin\&TRPM. SSEP\&TRPM achieves the best performance in minimum received energy at the expense of total received energy. Compared to the linear EH model of SSEP\&TRPM, the nonlinear EH model of SSEP\&TRPM is approximately 24\% higher at the minimum received energy and 6\% higher at the total energy, which shows that both performance improvements are achieved by considering the nonlinear EH model.

\begin{figure}[t]
\centering
\subfloat[]{
\includegraphics[width=0.32\linewidth]{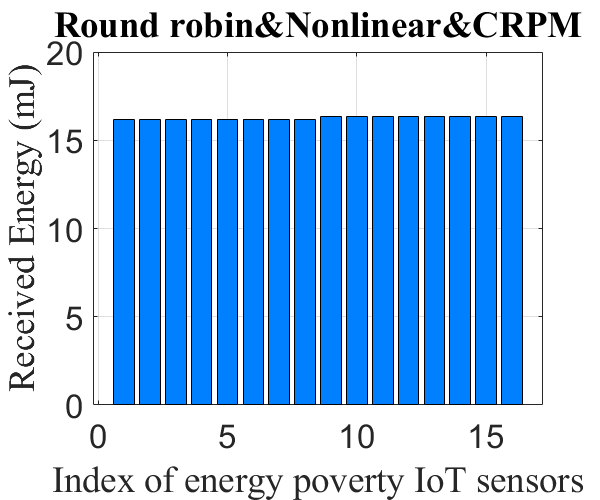}
}
\centering
\subfloat[]{
\includegraphics[width=0.32\linewidth]{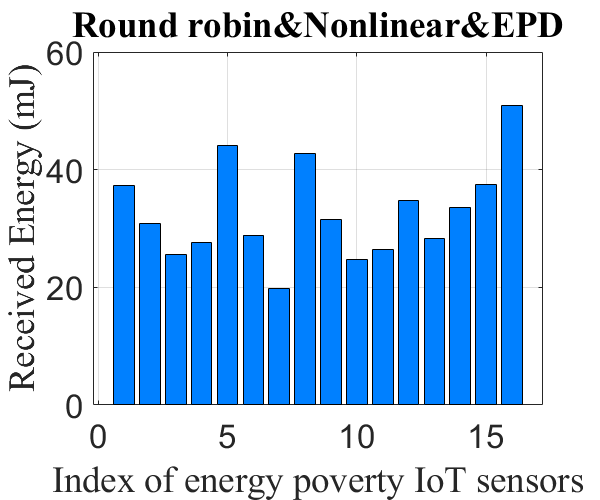}
}
\centering
\subfloat[]{
\includegraphics[width=0.32\linewidth]{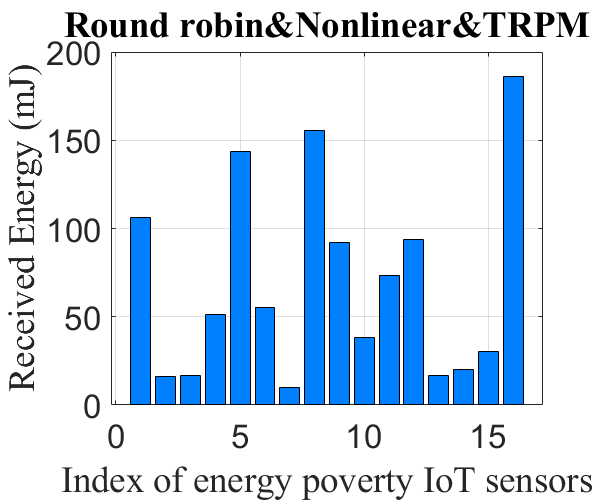}
}
\hskip2em
\centering
\subfloat[]{
\includegraphics[width=0.32\linewidth]{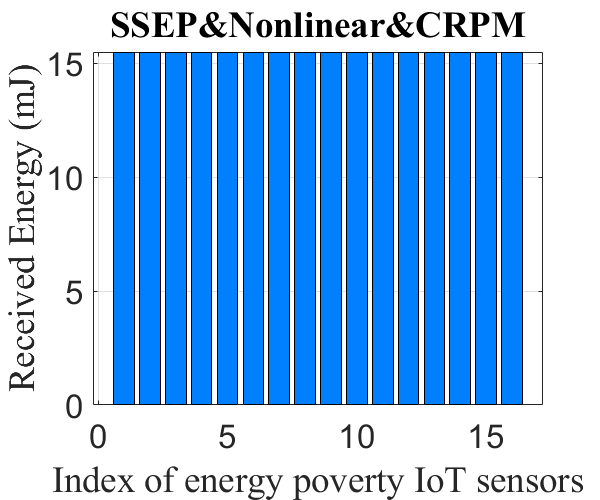}
}
\centering
\subfloat[]{
\includegraphics[width=0.32\linewidth]{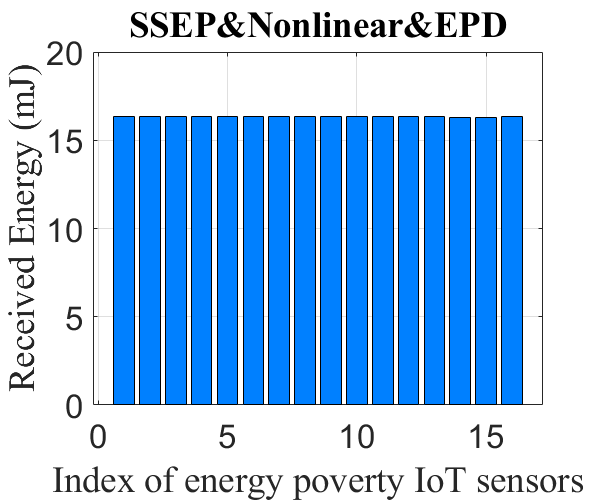}
}
\centering
\subfloat[]{
\includegraphics[width=0.32\linewidth]{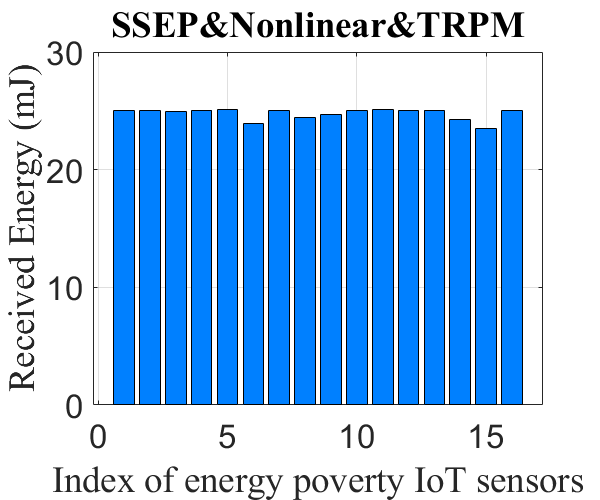}
}
\hskip2em
\centering
\subfloat[]{
\includegraphics[width=0.32\linewidth]{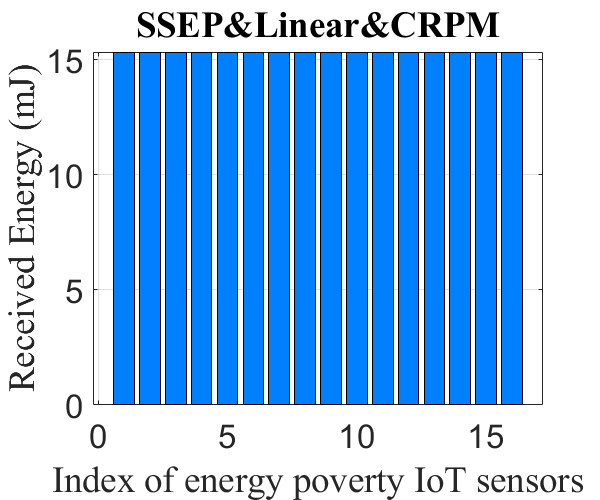}
}
\centering
\subfloat[]{
\includegraphics[width=0.32\linewidth]{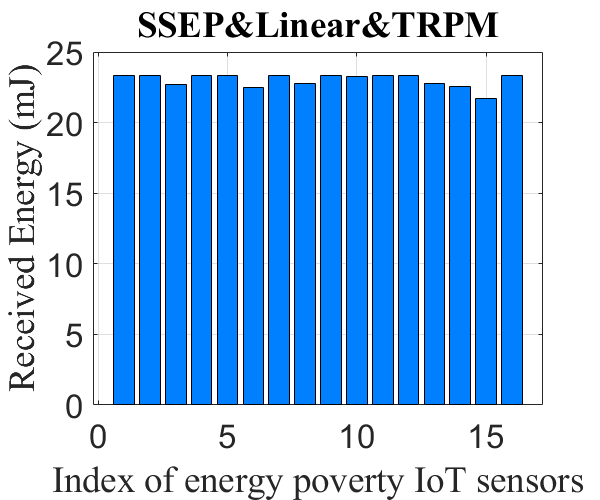}
}

\caption{Total received the energy of EPISs according to the resource allocation methods  with walk step: 0.2 m \\(a) Round-robin\&Nonlinear\&CRPM \\ (b)~Round-robin\&Nonlinear\&EPD \\ (c) Round-robin\&Nonlinear\&TRPM \\(d) SSEP\&Nonlinear\&CRPM\\ (e) SSEP\&Nonlinear\&EPD\\ (f) SSEP\&Nonlinear\&TRPM \\(g) SSEP\&Linear\&CRPM \\ (h) SSEP\&Linear\&TRPM
 }
\label{fig:HighPattern}
\end{figure}  

Fig.~\ref{fig:HighPattern} represents the total received energy among all EPISs with 0.2 m/iteration walk-step for the number of 10,000 power transmission iteration, under different transmission power and orthogonal bands allocation schemes. 
In SSEP, although some performance variations among EPISs appear when SSEP\&TRPM is used, it is relatively small than Round robin as shown in Fig.~\ref{fig:HighPattern}(c) and (f). Thus, we can verify the effectiveness in energy fairness issue of our proposed SSEP scheme, when providing power to EPISs with high mobility. 
In particular, when SSEP\&TRPM is used, EPISs have received similar energy between them, as opposed to 0.03 m/iteration walk-step. It indicates that if the EPIS's mobility is high, the EPISs with poor wireless link-status may move closer to PT sometimes, giving them chances for receiving a substantial amount of energy. Therefore, when SSEP\&TRPM is used, the EPISs with high mobility can receive energy in a balanced way, even in the case of TRPM.

\section{Conclusion} \label{conclusion}
We investigate a MISO-WPT system comprising of a multi-antenna power transmitter and multiple single-antenna EPISs. To handle energy fairness issues, we consider orthogonal bands allocation to the EPISs and energy beamforming technique on each orthogonal band. We propose orthogonal bands assignment rule based on the energy poverty of EPISs, granting the priority to the EPISs with less received energy. In addition, we formulate the common received power maximization (CRPM) problem that equalizes the received power of EPISs and the total received power maximization (TRPM) problem that maximizes the received power of EPISs. By considering the nonlinear EH model in optimization problems, it prevents the misleading optimized solution and reducing performance degradation. To solve the CRPM problem, the iterative bisection search method is adopted. For the sake of applying the bisection search method to the problem, this paper proposes a method of specifying the scope of the solution for the objective function defined by the sum of monotonous functions. A modified water-filling algorithm is applied to the closed-form solution obtained by using the KKT condition in the TRPM problem. Our extensive numerical results verify the importance of the proposed algorithms. Based on a comparison with Round robin scheduling algorithm, we validate the performance of our proposed SSEP scheme. In addition, we compare the proposed power allocation algorithms with an equal power distribution algorithm. Considering the mobility of EPISs as the one-dimensional random walk model, the effects of the mobility of EPISs on the minimum received energy of EPISs are presented. Unlike our belief, TRPM algorithms can achieve the best performance of both minimum and total received energy of all EPISs with high mobility. We also demonstrate the effectiveness of the nonlinear EH model by comparing the linear EH model, in TRPM and CRPM respectively.   
We conclude this paper with interesting future extensions of the presented schemes as follows:
 \begin{itemize}	
\item We considered the mobility of EPISs as a one-dimensional random walk model. In practice, EPISs may dynamically move around in the two-dimensional or three-dimensional space, such that it needs to consider a more complex mobility model of EPISs. 
\item We will extend the scenario from one power transmitter to multiple power transmitter. Multiple power transmitter may cooperate for energy beamforming, such that it is required that how to achieve the phase and frequency synchronization among multiple power tranmitter for energy beamforming.
\item It is interesting to extend the current design by combing the wireless communication system. The current work focuse on resource allocation on the WPT system. Simultaneous wireless information and power transmission system can be implemented in the proposed system to support various IoT services.  
\item Other practical setups in the system model can be considered, such as relay channel, imperfect channel state information, interference channel, etc.
\end{itemize}

\appendices
\section{Proof of Theorem~\ref{thm:TRPM} }

	\label{appendix:KKT}
This appendix shows the proof of Theorem~\ref{thm:TRPM}.	
Since the object function is concave and constraints are convex form,   we can apply KKT condition by reformulating this prblem as minimizing convex problem. Then, we can simplify the original problem as following equation using KKT condition.
	\begin{align}
		\label{KKT-1} \frac{a_kb_k\lambda_k}{1+b_k\lambda_kp^{t*}_{k}}+\mu_k-\zeta_k-\nu = 0,  \qquad\forall k \in \Gamma_P,    \\ 
		\label{KKT-2} \zeta_k\left(p^{t*}_{k}-\min\left(\frac{c_k}{\lambda_k},P_c\right)\right)=0, \qquad \forall k \in \Gamma_P,  \\
		\label{KKT-3} \mu_kp^{t*}_{k} = 0, \qquad\qquad\qquad \forall k \in \Gamma_P,  \\
		\label{KKT-4}\sum_{k \in \Gamma_P}p^{t*}_{k} - E_c = 0,  \quad \qquad \qquad\qquad\qquad \\
		\label{KKT-5} 0 \le p^{t*}_{k} \le \min\left(\frac{c_k}{\lambda_k},P_c\right),  \quad\qquad   \forall k \in \Gamma_P, \\
		\label{KKT-6} \mu_k \ge 0, \quad \zeta_k \ge 0,\quad \nu \ge 0, \qquad  \quad  \forall k \in \Gamma_P,  
	\end{align}  
	where $ \mu_k, \zeta_k,$ and $\nu$ is Largrange multipliers. There are three possible cases for $p_{t,k}$.  

\begin{itemize}
\item[1)]
  In first case, $0 < p^{t*}_{k} <  \min\left(\frac{c_k}{\lambda_k},P_c\right),$ it is clear that $ \mu_k =0 $ and $\zeta_k  = 0.$
 $\textrm{From the stationary condition (\ref{KKT-1})},$ we have $ p^{t*}_{k} = ha_k-\frac1{b_k\lambda_k} \quad \textrm{where}\quad  h= \frac1{\nu}.  $
\item[2)] In second case, $p^{t*}_{k}= 0$, the complementary slackness condition (\ref{KKT-2}) gives $\zeta_k = 0$ and from the stationary condition (\ref{KKT-1}), we have 
$h= \frac{1}{a_kb_k\lambda_k+\mu_k}$. It represents that $ha_k-\frac1{b_k\lambda_k}= \frac{a_k}{a_kb_k\lambda_k+\mu_k}-\frac{a_k}{a_kb_k\lambda_k}\le 0$. 
\item[3)] In third case, $p^{t*}_{k} = \min\left(\frac{c_k}{\lambda_k},P_c\right)$, we have $\mu_k =0$  from the complementary slackness condition (\ref{KKT-3}) 
and $h = \frac{ 1+b_k\lambda_k\min\left(\frac{c_k}{\lambda_k},P_c\right)}{a_k\left(b_k\lambda_k-\frac{\zeta}{a_k}\left(1+b_k\lambda_k \min\left(\frac{c_k}{\lambda_k},P_c\right)   \right)  \right)     } $ from the stationary condition (\ref{KKT-1}). Since $a_k, b_k, \lambda_k > 0$, we have $h>0$. 
By using the above cnoditions, it yields that\\ 
 $ha_k-\frac{1}{b_k\lambda_k} = \frac{1}{b_k\lambda_k-\frac{\zeta}{a_k}\left(1+b_k\lambda_k \min\left(\frac{c_k}{\lambda_k},P_c\right)   \right)} + \frac{b_k\lambda_k\min\left(\frac{c_k}{\lambda_k},P_c\right)}{b_k\lambda_k-\frac{\zeta}{a_k}\left(1+b_k\lambda_k \min\left(\frac{c_k}{\lambda_k},P_c\right)   \right)} \ge \min\left(\frac{c_k}{\lambda_k},P_c\right).  $ \\
Therefore, the optimal solution (\ref{TRPMsol}) is obtained by considering the above three cases and the primal feasibility condition (\ref{KKT-5}). 

\end{itemize}

	\section{Proof of Lemma~\ref{lemma:1}}   
	\label{appendix:SSSM}
	This appendix shows the proof of Lemma~\ref{lemma:1}. First, the $\alpha_{\text{min}} = \min_k a_k \log\left(1+ b_k\lambda_k\frac{E_c}{n(\Gamma_P)} \right) $ is obtained by substituting following $ p^{t}_{k} = \frac{E_c}{n(\Gamma_P)}$ and minizing the $\alpha$. Since $p^{t}_{k}(\alpha)$ is increasing and continuous function over $\alpha$, $p^{t}_{k}(\alpha_\text{min})$ has lower value than $\frac{E_c}{n(\Gamma_P)}$. Therefore, we can derive the following relationship:
	\begin{align*} 
		\sum_{k \in \Gamma_P} p^{t}_{k}(\alpha_{\text{min}}) &=& p^{t}_{1}(\alpha_{\text{min}}) + p^{t}_{2}(\alpha_{\text{min}})
		+ \dots + p^{t}_{n(\Gamma_P)}(\alpha_{\text{min}}) \\ 
		&\le& \frac{E_c}{n(\Gamma_P)} + \frac{E_c}{n(\Gamma_P)} + \dots + \frac{E_c}{n(\Gamma_P)} = E_c.  
	\end{align*} 
		Second, repeat the same process about $\alpha_{\text{max}}  = \max_k a_k \log\left(1+ b_k\lambda_k\frac{E_c}{n(\Gamma_P)} \right) $. 
	
	\begin{align*}
		\sum_{k \in \Gamma_P} p^{t}_{k}(\alpha_{\text{max}}) &=& p^{t}_{1}(\alpha_{\text{max}}) + p^{t}_{2}(\alpha_{\text{max}})
		+ \dots + p^{t}_{n(\Gamma_P)}(\alpha_{\text{max}}) \\ 
		&\ge& \frac{E_c}{n(\Gamma_P)} + \frac{E_c}{n(\Gamma_P)} + \dots + \frac{E_c}{n(\Gamma_P)} = E_c.  
	\end{align*}
	Since $p^{t}_{k}(\alpha_\text{max})$ has larger value than $\frac{E_c}{n(\Gamma_P)}$, the above relationship is satisfied. Therefore, $\alpha$ satisfying the condition (14) exists in the interval $\left[ \alpha_\text{min}, \alpha_\text{max}\right]. $

		\section{Pseudo-code for LTRPM}
		\label{APPENDIX LTRPM}
		This appendix shows the pseudo-code for the Linear Total Received Power Maximation algorithm. The algorithm is based on sequential resouce allocation algorithm.
	
	\begin{algorithm}[H]
		\caption{Linear Total Received Power Maximization Algorithm}
		\label{alg:LTRPM}
		\begin{algorithmic}[1]
			\STATE \textbf{Initialization} Arrange the set of EPISs $\Gamma_P$ by the value of $h_k\lambda_k$ in descending order, \\ 
			the remaining energy $E_r = E_c$, \\  
			the allocated power vector of EPISs $\boldsymbol{p^t} = 0$. \\ 
			\FOR { i = 1 : $|\Gamma_P|$ }
			\IF {  $ E_r > \min\left(\frac{c_k}{\lambda_k},P_c\right)$     }
			\STATE Set $p^{t}_{i} = \min\left(\frac{c_k}{\lambda_k},P_c\right).$ 
			\ELSE
			\STATE Set $p^{t}_{i} \leftarrow E_r$. 
			\ENDIF
			\ENDFOR
			\RETURN Set of allocated transmission power of EPISs $\boldsymbol{p^t}$ , after reaggraged by the original order.
		\end{algorithmic}
	\end{algorithm}

\section{Pseudo-code for LCRPM}
		\label{APPENDIX LCRPM}
			This appendix shows the pseudo-code for the Linear Commonl Received Power Maximation algorithm. The algorithm is based on iterative resouce allocation algorithm.

	\begin{algorithm}[H]
		\caption{Linear Common Received Power Maximization Algorithm}
		\label{alg:LCRPM}
		\begin{algorithmic}[1]
			\STATE \textbf{Initialization} The number of Iteration $I$= 0,\\
			the limit of Iteration $I^{\text{lim}}$, \\
			the remaining energy $E_r = E_c$, \\  
			the allocated power vector $\boldsymbol{p^t} = 0$, \\ 
			and the index set of EPISs $\Gamma_P.$ \\
			\WHILE {$\{E_r > 0\}$  and  $\{I < I^{\text{lim}}     \}  $ }
			\STATE Set $\alpha = \frac{E_r}{\sum_{k \in {\Gamma_P}}\frac{1}{h_k\lambda_k}}.$
			\FOR {$ k \in \Gamma_P$}
			\STATE Set $p^{t}_{k} \leftarrow p^{t}_{k} + \frac{\alpha}{h_k\lambda_k}. $
			\ENDFOR
			\STATE Set $E_r$ = 0.
			\FOR {$ k \in \Gamma_P$}
			\IF { $p^{t}_{k} > \min\left(\frac{c_k}{\lambda_k},P_c\right)   $ }
			\STATE Set $E_r = E_r + p^{t}_{k} - \min\left(\frac{c_k}{\lambda_k},P_c\right),   $\\
			$p^{t}_{k} = \min\left(\frac{c_k}{\lambda_k},P_c\right)$, \quad and \quad $\Gamma_P =  \Gamma_P \setminus k.$
			\ENDIF
			\ENDFOR
			\STATE $I = I+1.$ 
			\ENDWHILE
			\RETURN Allocated transmission power vector $\boldsymbol{p^t}$.
		\end{algorithmic}
	\end{algorithm}

\section*{Acknowledgment}
This work was supported by Institute of Information $\&$ communications Technology Planning $\&$ Evaluation (IITP) grant funded by the Korea government(MSIT) (2018-0-00691, Development of Autonomous Collaborative Swarm Intelligence Technologies for Disposable IoT Devices)

\ifCLASSOPTIONcaptionsoff
  \newpage
\fi

\bibliographystyle{IEEEtran}
\bibliography{ref}



\end{document}